\documentclass[useAMS,usenatbib]{mn2e}

\usepackage{graphicx,array}

\title[ISM properties in galaxy simulations]{ISM properties in hydrodynamic galaxy simulations:\\
Turbulence cascades, cloud formation, role of gravity and feedback}

\author[Bournaud et al.]{
Fr\'ed\'eric Bournaud$^{1}$\thanks{E-mail: frederic.bournaud@cea.fr}, Bruce G. Elmegreen$^{2}$, Romain Teyssier$^{1,3}$, David L. Block$^{4}$,\newauthor  Iv\^anio Puerari$^{5}$
\\
$^{1}$Laboratoire AIM Paris-Saclay, CEA/IRFU/SAp -- CNRS -- Universit\'e Paris Diderot, 91191 Gif-sur-Yvette Cedex, France.\\
$^{2}$IBM T. J. Watson Research Center, 1101 Kitchawan Road, Yorktown Heights, New York 10598 USA.\\
$^{3}$Institute for Theoretical Physics, University of Z\"urich, CH-8057 Z\"urich, Switzerland.\\
$^{4}$School of Computational and Applied Mathematics, University of the Witwatersrand, Private Bag 3, WITS 2050, South Africa.\\
$^{5}$Instituto Nacional de Astrof\'\i sica, Optica y Electr\'onica, Calle Luis Enrique Erro 1, 72840 Santa Mar\'\i a Tonantzintla, Puebla, Mexico.
}
\begin{document}

\date{Accepted 2010 July 14 --- Received 2010 May 8 --- in original form 2010 May 8}


\maketitle

\label{firstpage}

\begin{abstract}
We study the properties of ISM substructure and turbulence in hydrodynamic (AMR) galaxy simulations with resolutions up to 0.8~pc and $5\times 10^3$~M$_{\sun}$. We analyse the power spectrum of the density distribution, and various components of the velocity field. We show that the disk thickness is about the average Jeans scale length, and is mainly regulated by gravitational instabilities. From this scale of energy injection, a turbulence cascade towards small-scale is observed, with almost isotropic small-scale motions. On scales larger than the disk thickness, density waves are observed, but there is also a full range of substructures with chaotic and strongly non-isotropic gas velocity dispersions. The power spectrum of vorticity in an LMC-sized model suggests that an inverse cascade of turbulence might be present, although energy input over a wide range of scales in the coupled gaseous+stellar fluid could also explain this quasi-2D regime on scales larger than the disk scale height. Similar regimes of gas turbulence are also found in massive high-redshift disks with high gas fractions. Disk properties and ISM turbulence appear to be mainly regulated by gravitational processes, both on large scales and inside dense clouds. Star formation feedback is however essential to maintain the ISM in a steady state by balancing a systematic gas dissipation into dense and small clumps. Our galaxy simulations employ a thermal model based on a barotropic Equation of State (EoS) aimed at modelling the equilibrium of gas between various heating and cooling processes. Denser gas is typically colder in this approach, which is shown to correctly reproduce the density structures of a star-forming, turbulent, unstable and cloudy ISM down to scales of a few parsecs.
\end{abstract}

\begin{keywords}
galaxies: evolution -- galaxies: ISM -- galaxies: star formation -- galaxies: structure -- ISM: kinematics and dynamics -- ISM: structure\end{keywords}

\section{Introduction}

Observations of the interstellar medium (ISM) of nearby galaxies reveal a large variety of complex substructures, such as density waves, molecular clouds, filamentary structures along shocks, and supernova bubbles. ISM turbulence is a major driver of large-scale star formation: turbulent compression can increase the gas density and initiate star-forming instabilities \citep{E93,E02}. This can happen in the spiral arms of modern disk galaxies \citep{klessen2001}, and there is recent evidence that very active star formation relates to increased turbulence and associated instabilities in both primordial disk galaxies \citep{E05,FS06,BEE07,DSC09} and starbursting mergers \citep{bournaud08a,teyssier10}. On the other hand, gas turbulence can also disrupt gaseous structures faster than they gravitationally collapse, which could potentially stabilize gas disks and quench the star formation activity of some galaxies \citep{martig09,DSC09}.

Nevertheless, the large-scale properties of ISM turbulence are still poorly understood and the ability of numerical simulations to realistically model ISM substructures remains largely unexplored. Hydrodynamic simulations of ISM turbulence and fragmentation in whole galaxy models have been performed \citep{wada02,wada07,tasker08,tasker09,agertz09} but have not been compared to observational constraints yet, probably because of lacking the required level of realism, both in spatial resolution and in global dynamical modeling (no old stellar components and live dark halo were considered), leading to a restricted dynamical range of structure sizes. All these models are characterized by a probability distribution function (PDF) of the gas density that has a log-normal shape, resulting from the combination of various structures including density waves, dense gas clouds, supernovae bubbles. The levels of turbulent speed in these models seem realistic for both nearby and high-redshift galaxies \citep[see][respectively]{agertz09,ceverino10}. Whether a realistic spectrum of structures of various scales is reproduced in modern hydrodynamic models remains however unknown.

Observations suggest that turbulent motions are initiated by gravitational instabilities on a relatively large scale (the Jeans scale, which is $1/k \sim 100$~pc for wavenumber $k$ in nearby disk galaxies, Elmegreen et al. 2003) and that a turbulent cascade develops towards smaller scales. This turbulence should be three-dimensional since the Jeans scale length is expected to set the gas disk scale height. Turbulent motions could be triggered by gravity even inside molecular clouds \citep{field}.

Observations of the power spectra of ISM emission from nearby galaxies are consistent with this. On scales smaller than $\sim100$ pc, the power spectrum is steep, $\sim-3$, as it is for 3D Kolomogorov turbulence, while on scales larger than $\sim100$ pc, the power spectrum is flatter by about 1 in the slope, with a power-law exponent of $\sim-2$.
This has been observed in the Large Magellanic Cloud (LMC) (Elmegreen, Kim, \& Staveley-Smith 2001; Block et al. 2010), and in NGC 1058 (Dutta et al. 2009).
This tentatively suggests a two-dimentional regime for turbulent motions on large scales, because the break in the density distribution occurs at the expected disk scale height. The nature of the large-scale motions is uncertain, however, because the three-dimensional velocity field in individual galaxies is not observable. In the Small Magellanic Cloud, the power spectrum is steep everywhere (Stanimirovic et al. 2000), presumably because there is no thin disk on the line of sight.

In this paper, we explore the properties of ISM turbulence in simulations of isolated disk galaxies, performed with an adaptive grid hydrodynamic code (AMR) with a maximal resolution of 0.8~pc resolution. We use a ``pseudo-cooling'' approach to model the equilibrium of gas between various heating and cooling processes, based on a barotropic equation of state (EoS), and perform models with and without supernovae feedback. Our main model has a mass distribution and rotation curve representative for the LMC disk, so as to compare with observations from Block et al. (2010). However we do not try to reproduce the detailed individual structures of the real LMC, and we also analyze models with higher disk masses and gas fractions. We show that a large range of substructures spontaneously form in the ISM, with a power spectrum quite consistent with observations of the LMC from very large scales ($> 1$~kpc) down to very small scales ($< 10$~pc). A break in the power-law density power spectrum is observed at the gas disk scale height. Studying the velocity structure on various scales, we show that this corresponds to a transition from a 3D regime to a (quasi-)2D regime in the gas motion, as initially proposed from observations (Elmegreen et al. 2001). Even if large-scale forcings are probably important, the two-dimensional regime on large scales cannot be interpreted just as density waves and streaming motions in a rotating disk, and evidence for an inverse cascade of turbulence in a quasi-2D disk is proposed. The same properties are recovered in a model of a massive high-redshift disk galaxy, which has a high gas fraction (50\%) and strong turbulence ($V/\sigma \sim 4$), suggesting these properties are relatively universal for gaseous galactic disks.

Analyzing the vertical structure of the disk, compared to the properties of gas turbulence, we suggest that the turbulent motions at all scales are mainly powered by gravitational instabilities around the Jeans scale length, which sets the disk scale height. Feedback processes such as supernovae explosions do not significantly change the initial statistical properties of the ISM. They are nevertheless fundamental to maintain a realistic ISM structure over a large number of disk rotations. Supernovae bubbles do not appear to set the disk scale height, but they expand up to the disk scale height that is maintained by gravitationally-driven turbulence: this breaks apart the densest self-gravitating clouds to maintain a steady state cloud population.

\section{Simulations}

\begin{figure}
\centering
\includegraphics[width=7.5cm]{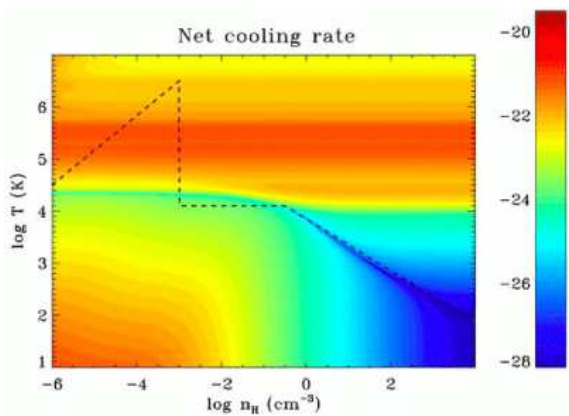}
\caption{The ``pseudo-cooling'' Equation of State used in our simulations. Gas cooling and heating rates (in erg~s$^{-1}$~cm$^{-3}$) as a function of temperature and density is shown for 1/3 solar metallicity, as computed with the CLOUDY code \citep[][see text]{ferland}. The EoS used in our simulations is shown by the 
dashed line. A low-density regime corresponds to the Virial temperature of hot gas in a diffuse halo. Denser gas follows an equilibrium given by the minimal cooling rate: an isothermal regime accounts for warm HI in the disc at $10^4$~K, and denser phases corresponding to cool/cold atomic and molecular gas. The EoS follows the thermal equilibrium, and variations of metallicity and UV background do not largely shift this equilibrium, so the same EoS would remain sensible for higher metallicities or UV fluxes. A thermal support at very high density is added to this EoS to prevent artificial fragmentation (see text).}
\label{fig:eos}
\end{figure}

\begin{figure}
\centering
\includegraphics[width=8cm]{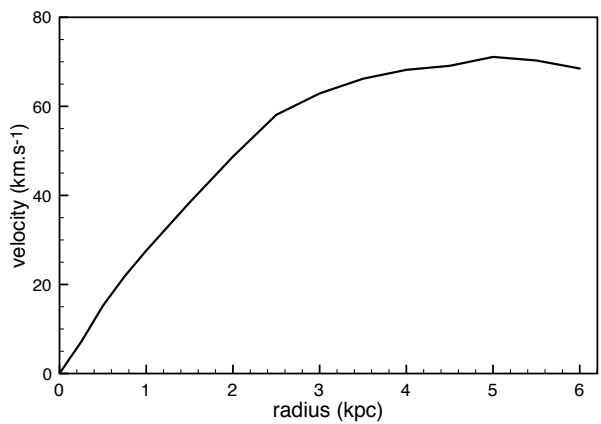}
\caption{Rotation curve of our LMC-sized model, measured at $t=254$~Myr.}
\label{fig:model-rc}
\end{figure}

\subsection{Simulation technique}

We perform closed-box model of isolated galaxies, with physical parameters described in section~\ref{model:params}. We use the AMR code RAMSES \citep{teyssier2002}. Stars and dark matter are described with collisionless particles, evolved with a particle-mesh (PM) technique. The gaseous component is described on the same adaptive grid; hydrodynamic equations being solved with a second-order Godunov scheme.

\begin{figure}
\centering
\includegraphics[angle=270,width=8cm]{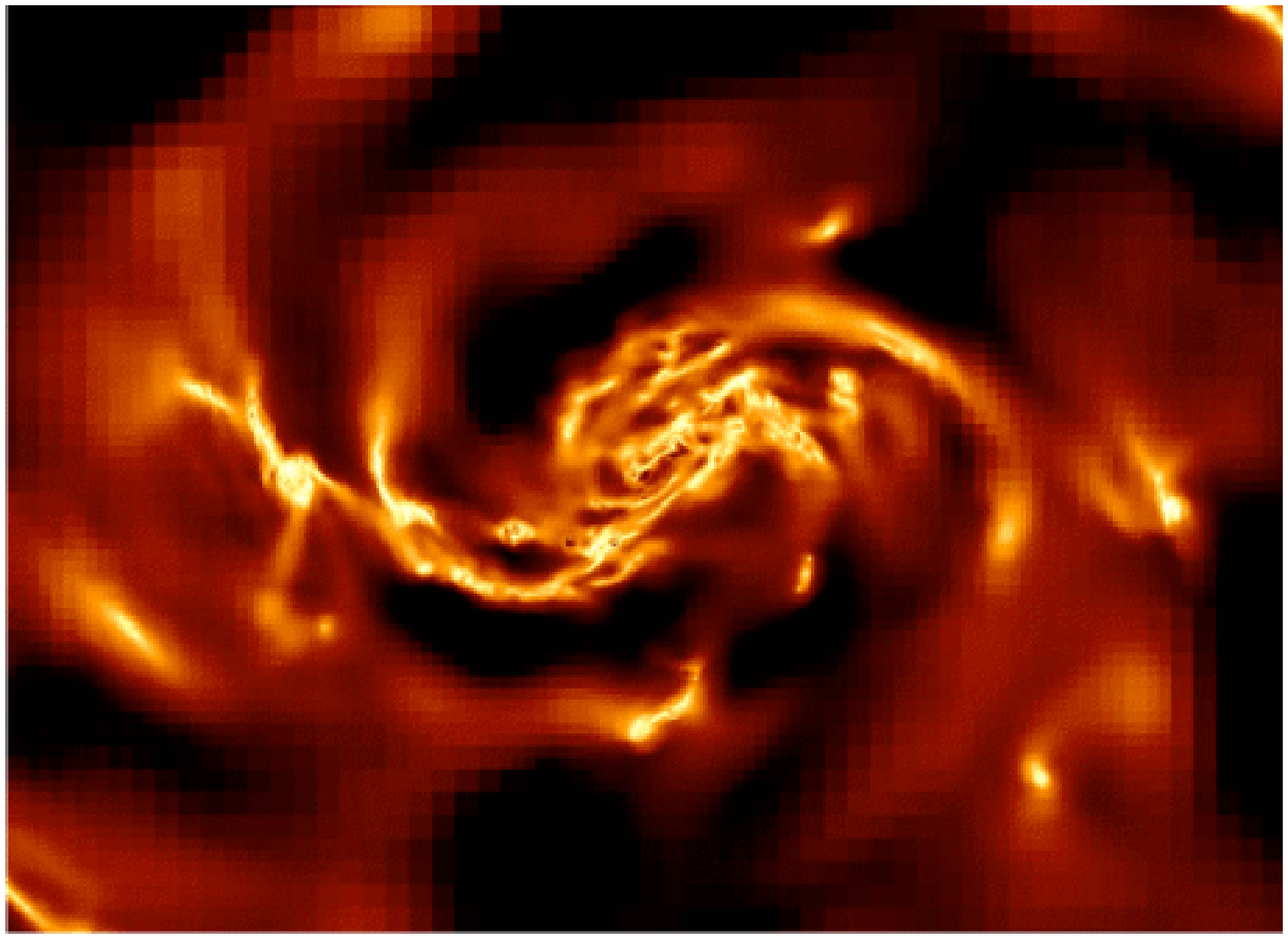}\\
\includegraphics[angle=270,width=8cm]{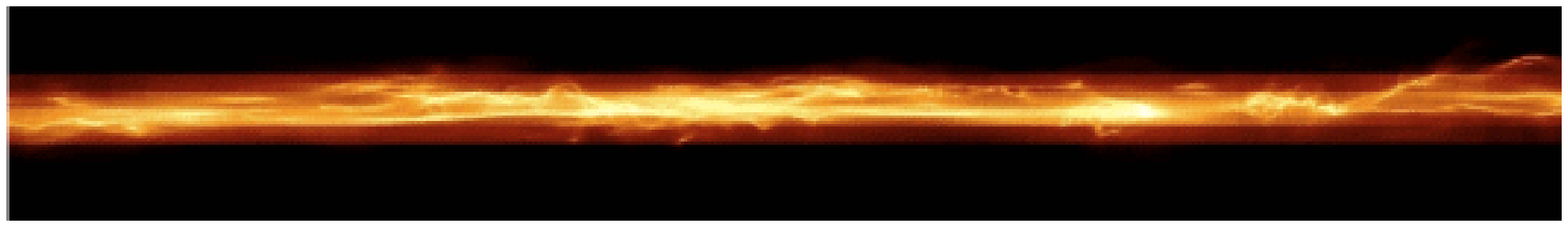}\\
\includegraphics[angle=270,width=8cm]{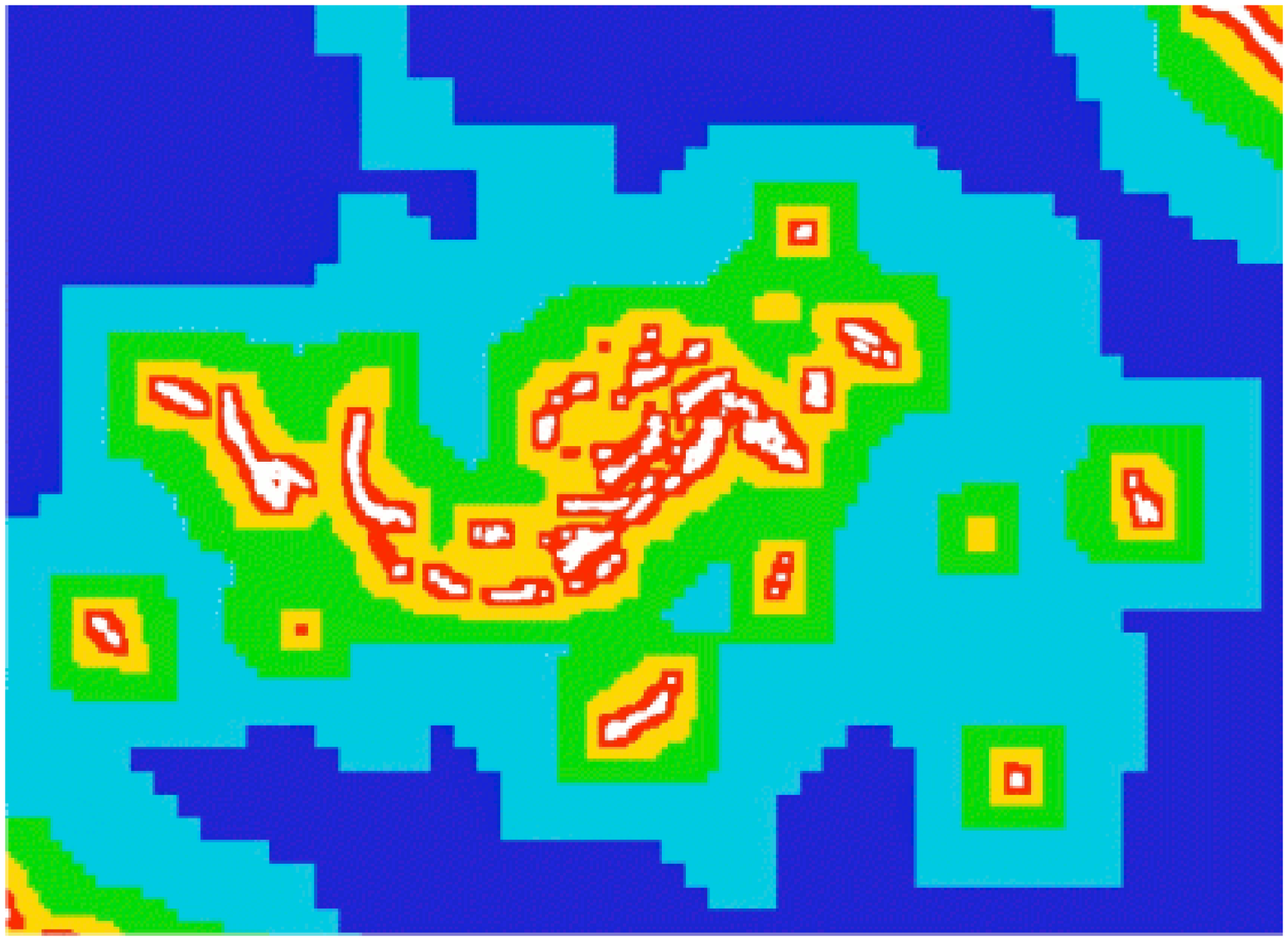}
\caption{{\bf Top:} Gas density maps of the LMC-sized gas disk with feedback, at $t=254$~Myr, in a face-on projection ($7\times4$~kpc snapshot) and a side-on projection ($7\times 0.8$~kpc snaphot). {\bf Bottom:} Map of the AMR grid refinement level in the mid-plane, with $l=15$ in white and $l=11$ in dark blue.}
\label{fig:sim-density}
\end{figure}

Our simulations use a box size of 26~kpc. The coarse level of the AMR grid, $l_{min}=9$, corresponds to a $(2^9)^3 = 512^3$ Cartesian grid, i.e. the minimal spatial resolution is $\epsilon_{min} = 51$~pc where $\epsilon = \frac{26}{l}$~kpc is the cell size. The maximal refinement level in dense regions is set by $l_{max}=15$, i.e. an effective resolution equivalent to $32768^3$ and a maximal spatial resolution of $\epsilon_{max} = 0.8$~pc. As we want to resolve turbulent motions in the whole disk, we use a refinement strategy that ensures that the mid-plane of the initial gas disk is entirely resolved with $l>11$ ($\epsilon < 13$pc), and with $l=12$ ($\epsilon=6$~pc) at the initial average surface density. The coarsest resolution levels are used only in the outer stellar bulge and halo. The resolution increases towards a cell size of 0.8~pc in the inner disk and in dense substructures forming in the outer disk. Effective resolution maps are shown on Figure~\ref{fig:sim-density}. The refinement is as follows: a cell of level $l<l_{max}$ is refined if its gas mass is larger than $m_{res}=5 \times 10^3$~M$_{\sun}$ and/or the number of particles (stars and dark matter) is larger than $n_{res}=8$.

Only the gas density is computed in the smallest cells of the AMR grid with $l=l_{max}=15$, the mass density of particles being restricted to $l_{max,part}=14$ in the PM scheme. This ensure a gravitational softening of at least 2~pc for particles. Treating the particle density at the $l_{max}$ level could result in a low number of particles per cell, and induce two-body relaxation, in regions that are refined because of a high gas density without necessarily having a high density of stars and/or dark matter. Only a few $l_{max}$ cells at each instant would actually contain less than 3 particles per cell, as measured in our simulation results, but we prefer avoiding any spurious effect. 

\smallskip

Resolving the gas cooling and heating equation is numerically costly and would reduce the achievable resolution. In order to resolve small-scale structures and gas turbulence at the best possible resolution, we model the cooling and heating processes using an EoS that fixes the gas temperature $T$ as a function of its local atomic density $n$. Here we use an EoS modeling the equilibrium of gas between the main heating (UV heating and stellar feedback) and cooling (atomic and dust cooling) processes \citep[see also][]{teyssier10}. This EoS is a fit to the average equilibrium temperature of gas at one third solar metallicity: for densities $10^{-3}<n<0.3$~cm$^{-3}$, $T(n)=10^4$~K. For densities $n>0.3$~cm$^{-3}$, $T(n)=10^4 (n/0.3)^{-1/2}$~K. Densities $n<10^{-3}$~cm$^{-3}$ correspond to gas in gravo-thermal equilibrium in the halo, at the Virial temperature $T(n)=4\times 10^6 (n/{10^{-3}})^{2/3}$~K.

Gravitational instabilities are a major driver of turbulence in galactic disks, both by directly triggering gas motions and by fueling star formation that can also pump turbulent energy through feedback processes. This is known from observations \citep{E02}, and models similar to ours but scaled to massive spiral galaxies \citep{wada02, tasker09, agertz09}. A fundamental aspect in our models is thus to avoid numerical instabilities and artificial fragmentation: a gas medium that should theoretically be stable, tends to be unstable in numerical models when the Jeans length is described by a small number of spatial resolution elements \citep{truelove97}. It is generally admitted that this effect is avoided if the Jeans length is resolved by at least a few elements \citep[e.g.]{machacek,agertz09}. We then force the grid refinement to ensure that the thermal Jeans length of the gas is always resolved by at least 4 cells -- and hence the actual thermal+turbulent Jeans length is resolved by an even larger number of cells). At the $l_{max}$ grid level,  we impose a temperature threshold computed to keep the thermal Jeans length larger than $4 \times \epsilon_{max}$. This temperature floor, added to the EoS at very high density ($\sim 10^{5}$~M$_{\sun}$~pc$^{-3}$ and above) can be considered as a sub-grid model for the unresolved turbulent motions at scales smaller than the maximal grid resolution. We can then consider that artificial fragmentation is generally avoided: simulations of galactic disk fragmentation imposing higher numbers of resolution elements per Jeans length do not show significant differences in the properties of gas stability and turbulence \citep[e.g.,][]{ceverino10}.

 \smallskip
 
 One of our models includes only disk self-gravity, but the other model also includes star formation and feedback. A threshold for star formation is implemented at a gas mass density $\rho = 5000$~atoms~cm$^{-3}$ $\simeq 115$~M$_{\sun}$~pc$^{-3}$. This choice ensures that stars form only in the densest gas structures formed in our models but that the density PDF remains correctly sampled around and above this threshold (see Figure~\ref{fig:sim-pdf}).
Above the threshold, the specific star formation rate (SFR per gas mass unit) is defined as the product of the efficiency and the local free-fall time$\sqrt{\frac{3 \pi}{32 G \rho}}$. We use an efficiency of 10\%, which was chosen to give a gas consumption timescale of 2~Gyr on the results of the first simulation without star formation. Note that lower resolution models usually need to combine a lower efficiency with a lower density threshold for star formation (see discussion in Teyssier et al. 2010).

We use the kinetic feedback model described by \citep{dubois}. A 20\% fraction of the mass of stars formed at a given timestep is assumed to be in stars that will explode into supernovae. 15\% of the typical energy of each supernova, $10^{51}$~erg, is re-injected into the gas, assuming that the remaining energy is radiated away. The energy is re-injected in the form of radial velocity kicks around the supernova, in a gas bubble of radius 3~pc (see Dubois \& Teyssier 2008 for details). 

We use outflow boundary conditions for the fluid, and isolated boundary conditions for gravity

\subsection{Model Parameters}\label{model:params}

Our simulations are not aimed at reproducing the exact morphology of the LMC, but at studying the properties of turbulence in a galactic disk with general properties similar to the LMC. To this aim, we initialize an exponential stellar disk containing $4 \times 10^6$ particles, of total mass $3\times 10^9$~M$_{\sun}$ radial scale-length of 1.5~kpc, and truncation radius of 3~kpc. A non-rotating bulge of $3\times 10^8$~M$_{\sun}$ with radius 500~pc, made-up of $4 \times 10^5$ particles, is added. The stellar velocity dispersions initially correspond to a stellar Toomre parameter of $Q_{\mathrm S}=1.5$, so that stars are stable against axisymmetric instabilities but unstable to bar forming instabilities -- a bar and a pair of spiral arms form spontaneously in our model.

The dark matter halo contains $4 \times 10^6$ particles, has a total mass of $5\times 10^9$~M$_{\sun}$ and follows a Burkert profile of core radius 3.5~kpc, truncated at 10~kpc. The initial gas disk is exponential, with a scale-length of 3~kpc and truncation radius of 5~kpc, an exponential scale-height of 50~pc truncated at 500~pc, and a gas mass of $6\times 10^8$~M$_{\sun}$. The gas disk is initially purely rotating, with no macroscopic velocity dispersion. The initial densities in the disk lie within the isothermal part of our EoS, so a thermal support of $\sim 10$~km~s$^{-1}$ ($T=10^4$~K) is present all over the initial disk, which can be considered as a model for the turbulent support of typically a few km~s$^{-1}$ in real disks. This support ensures an initial Toomre parameter for the gas $Q_{\mathrm G}\simeq 1-1.5$. Once dense structures form in the course of the simulation, the temperature decreases (following the EoS) and the thermal support is gradually replaced by a turbulent support.
Outside the initial disk truncation, the AMR grid is initialized with a background density of $10^{-6}$ atom per cm$^3$. 

The rotation curve in this model is shown on Figure~\ref{fig:model-rc}. It is broadly similar to the LMC rotation curve (Kim et al. 1998), with almost solid body rotation up to radii of at least 2~kpc, and gradually flattening with a circular velocity $\simeq 70$~km~s$^{-1}$ at large radii. We study the internal dynamics of the LMC; the interaction with the Milky Way and its hot gas halo is not modelled. Ram pressure was found to compress the leading edge of the HI and H$\alpha$ disk, with little effect on the inner star-forming disk (Mastropietro et al. 2009; see also Tonnesen \& Bryan 2009 on the effect of ram pressure on a cloudy ISM).

\begin{figure*}
\centering
\includegraphics[angle=270,width=17cm]{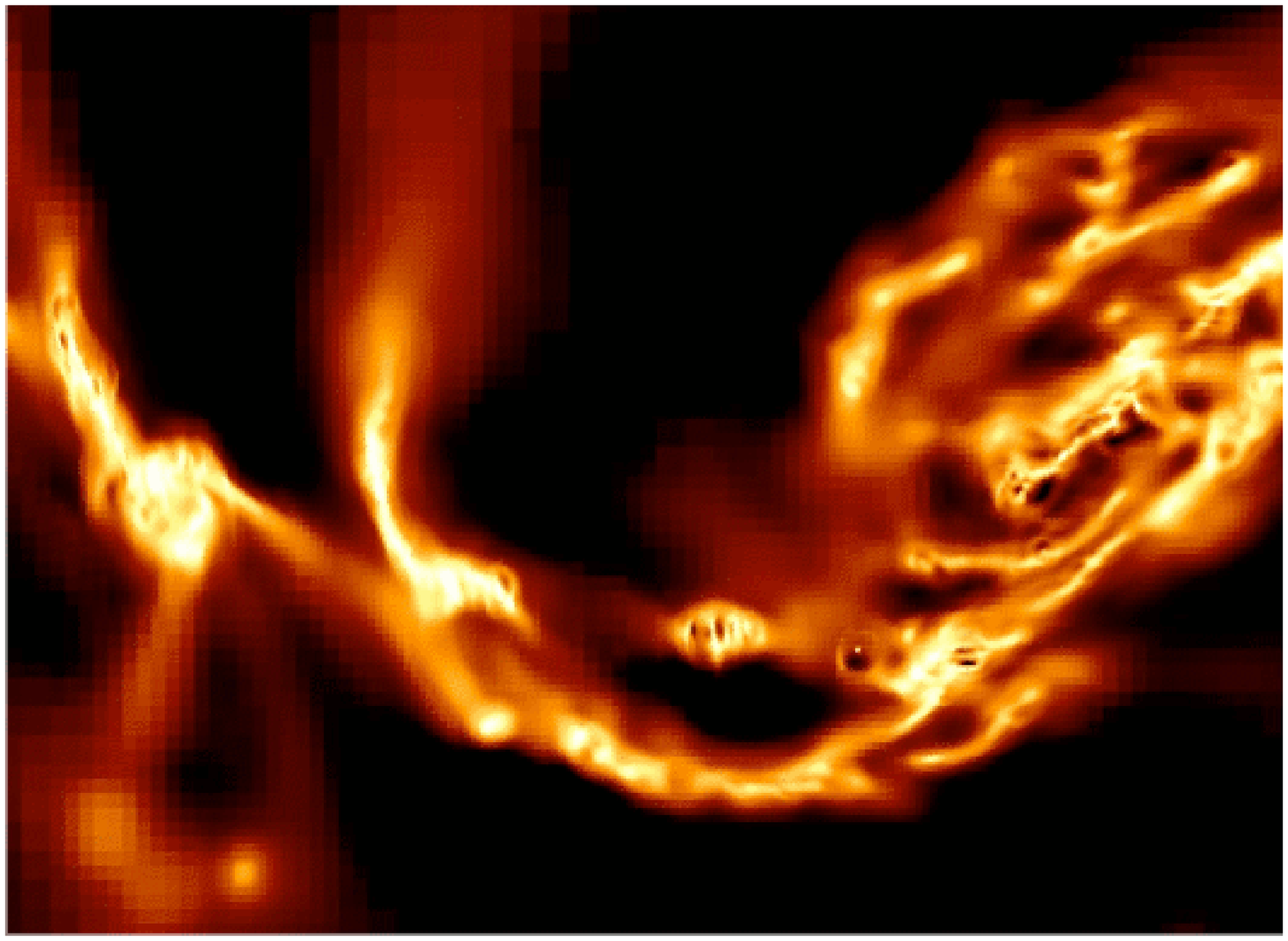}\\
\includegraphics[angle=270,width=17cm]{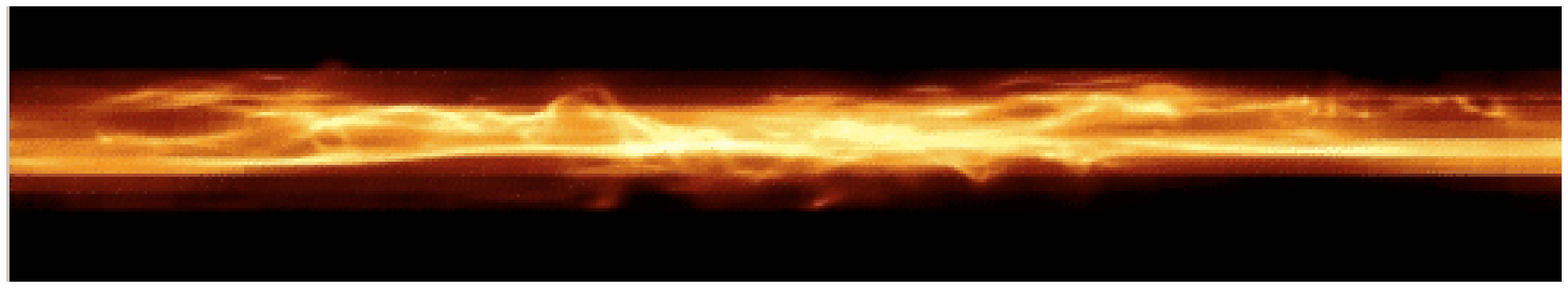}
\caption{Zoomed-in face-on and side-on density maps of the LMC-sized model with stellar feedback at $t=254$~Myr. The horizontal size of these figures is 3~kpc. The darkest regions visible on these projections have surface densities of 35~M$_{\sun}$~pc$^{-2}$: a low density fountain above the disk plane is not shown on the edge-on projected.} 
\label{fig:sim-density-zoom}
\end{figure*}

\begin{figure}
\centering
\includegraphics[width=8cm]{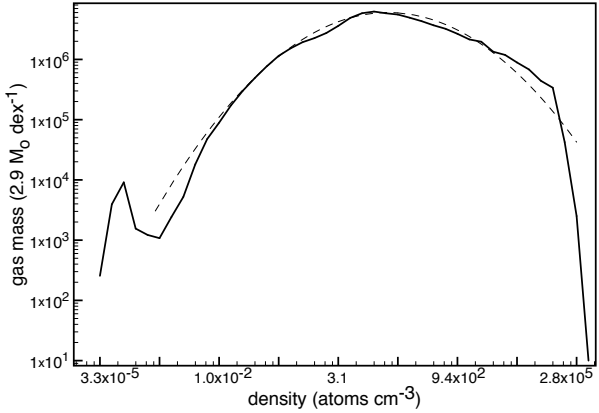}
\caption{Probability distribution function (PDF) of the gas density $\rho$ at $t=254$~Myr in the model with feedback. The dashed curve is a log normal profile centered on $\rho_\mathrm{max}=53$~cm$^{-3}$ and of standard deviation $\Delta$=1.27 in a base-10 log PDF (see text for details). The dotted line is for the simulation with reduced resolution (section~\ref{resol}). This PDF was measured within a cylindrical box of radius 5~kpc and height 2~kpc. The peak at low densities corresponds to low density gas in the hot halo.}
\label{fig:sim-pdf}
\end{figure}

\begin{figure}
\centering
\includegraphics[angle=270,width=8cm]{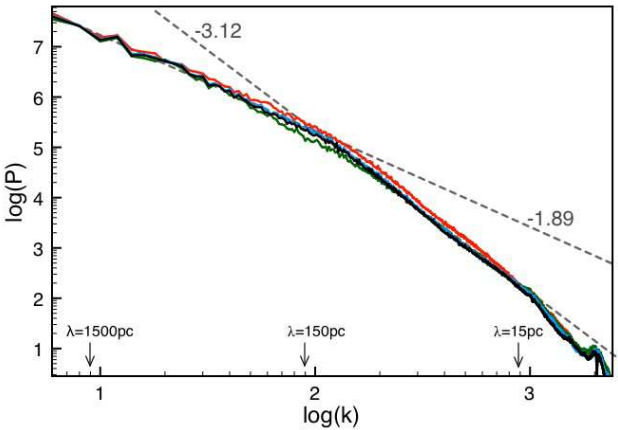}
\caption{Power spectrum of the face-on gas surface density in the model with feedback, at $t=254$~Myr (dark), 261~Myr (green), 268~Myr (blue) and 275~Myr (red). The wavenumber unit is $4.70 \times 10^{-4}$~pc$^{-1}$.} 
\label{fig:sim-powspec}
\end{figure}

\begin{figure}
\centering
\includegraphics[angle=270,width=8cm]{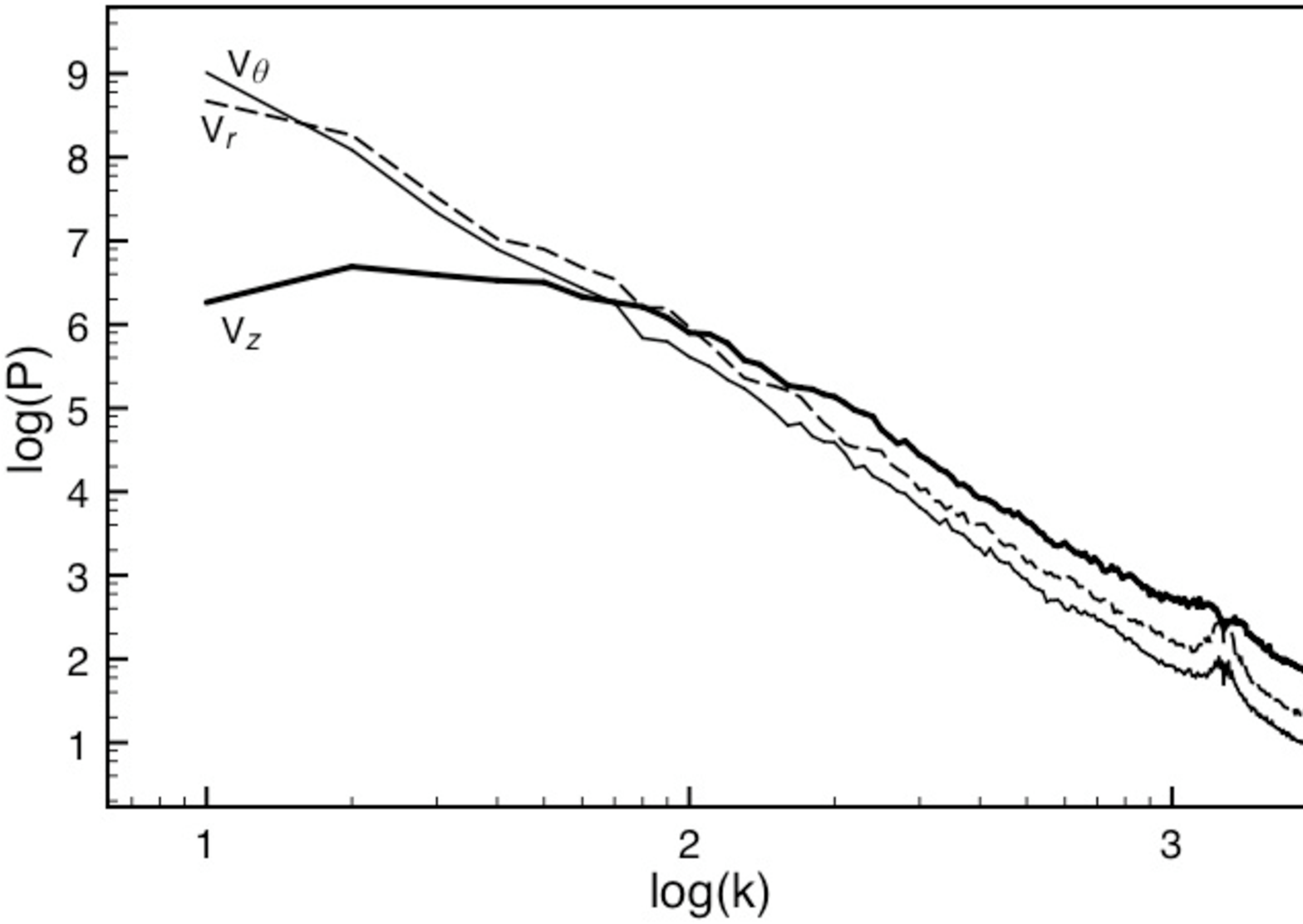}
\caption{Velocity power spectrum for the three separate components $V_r$, $V_\theta$ and $V_z$, at $t=254$~Myr (LMC-sized model with stellar feedback). The conversion of wavenumbers into linear size is as on Fig.~\ref{fig:sim-powspec}.}
\label{fig:sim-velspec}
\end{figure}

\newpage

\section{Results}

The main simulation analyzed in Sections 3.1 and 3.2 is the LMC-sized model with stellar feedback. We later compare to the simulation without star formation and feedback in Section~3.3, to better highlight the role of feedback in the global structure of the ISM.

\subsection{ISM structure and density power spectrum}

We measured the velocity dispersion $\sigma$ in the gas disk, and found that both the mass-weighted average value $<\sigma>$ and the standard deviation $\Delta \sigma$ rapidly increase in the first 200~Myr and do not significantly evolve after $t\sim 200$~Myr, indicating that a steady state in the turbulent regime has been reached. We thus mostly analyze our models around $t$=250~Myr. 

Spectral analysis of the gas distribution was performed on a sub-box of size 13~kpc (the outer regions of the full simulation box contain only the outer regions of the dark matter halo). Throughout the paper, the wavenumber $k=1$ is for a wavelength of 13~kpc, which results in a wavenumber unit $u_k = 4.70 \times 10^{-4}$~pc$^{-1}$.

\medskip

Figure \ref{fig:sim-density} shows the gas density distribution for this simulation at $t=254$~Myr. The response to a stellar bar, and some long spiral arms are seen, together with more chaotic structures such as shorter arms, shocks, dense clouds, bubbles, etc (Fig.~\ref{fig:sim-density-zoom}). We fit a sech$^2$ vertical profile to the gas at radii smaller than 5~kpc, integrated over the plane, and find a scale height of 207~pc. The scale height is relatively constant, with moderate flaring, and varies from 187~pc inside the central kpc to 226~pc for $3<r<5$~kpc. 

The density PDF, integrated over the central disk with a radius of 5~kpc and height of 2~kpc, is shown on Figure~\ref{fig:sim-pdf}. This PDF has a log-normal shape, truncated at quite high density ($\sim 10^6$~cm$^{-3}$) which is enabled by our high spatial and mass resolution. If we write the probability distribution for the density $$P(\rho) \propto e^{-0.5 \frac{ \left(\ln \frac{\rho}{\rho_{\mathrm max}} \right) ^2  }{\Delta^2}} \; ,$$ then the Gaussian dispersion is $\Delta \simeq 2.92$.

\medskip

The power spectrum of the face-on gas column density is shown on Figure~\ref{fig:sim-powspec} at the same instant and later ones. No significant time evolution is found. The power spectrum shows a double power-law with a break at wavelengths of about 150~pc, which is about the measured gas disk scale-height. The power law for large structures has a slope of $\simeq -1.9$ and the slope for small structures is $\simeq -3.1$. The slope steepens for the smallest scales below 5-10~pc, which is discussed later in section~3.3.

The simulation without star formation and feedback shows a globally similar density power spectrum (see Sect.~3.3 and Fig.~14) with a double power-law shape with slopes of $-2.9$ on small scales and $-1.8$ on large scales. The transition occurring around the gas disk scale-height. Only the smallest scales below 10~pc significantly differ from the model with star formation and feedback (see section~3.3).

\medskip

Our simulations produce a density structure that has about the power spectrum observed for the ISM in the LMC and other nearby galaxies. The power spectrum shows a break between a $-2$ power law and a $-3$ power law, with the transition occurring at scales around the gas disk scale-height, as usually speculated in observations \citep[e.g.,][]{E01,dutta}. In the following, we study the velocity structure to understand the nature of the motions in the two regimes of the density distribution.

\subsection{Velocity structure}

\begin{figure*}
\centering
\begin{tabular}{>{\centering}m{6.1cm} >{\centering}m{1.25cm} >{\centering}m{6.1cm}}
$V_r$ & & $V_z$
\tabularnewline
\includegraphics[angle=270,width=6cm]{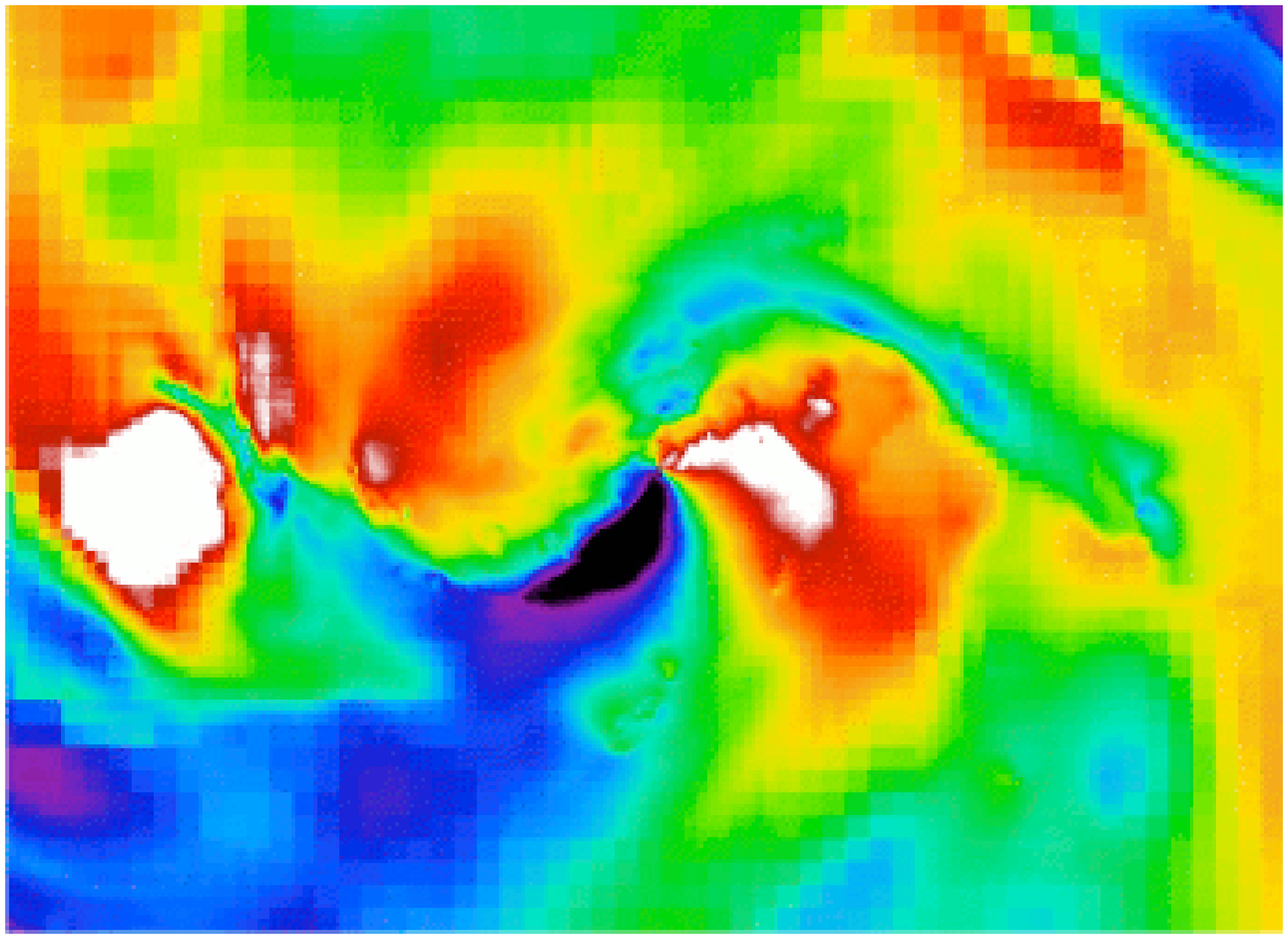} & total & \includegraphics[angle=270,width=6cm]{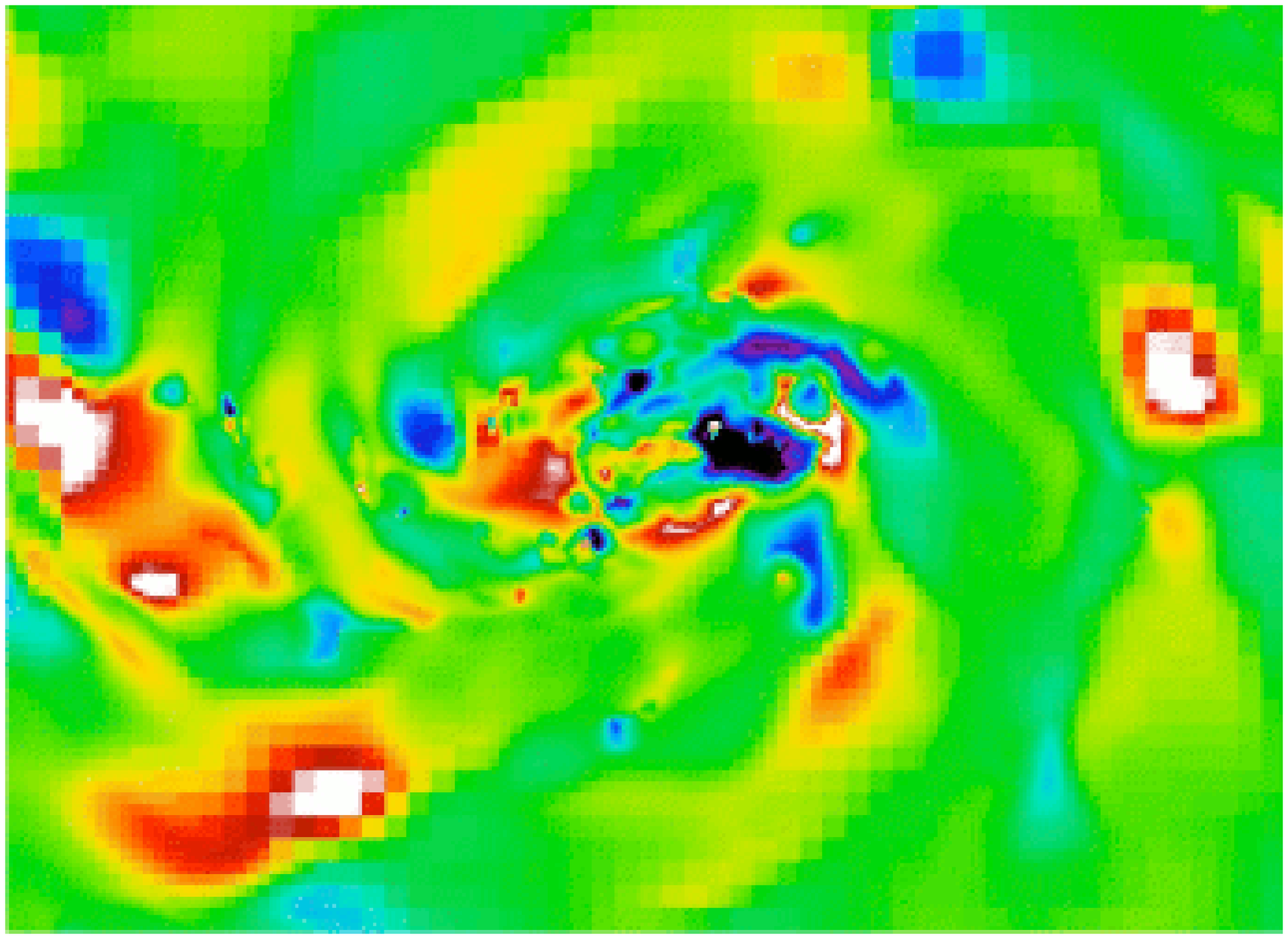} 
\tabularnewline
\includegraphics[angle=270,width=6cm]{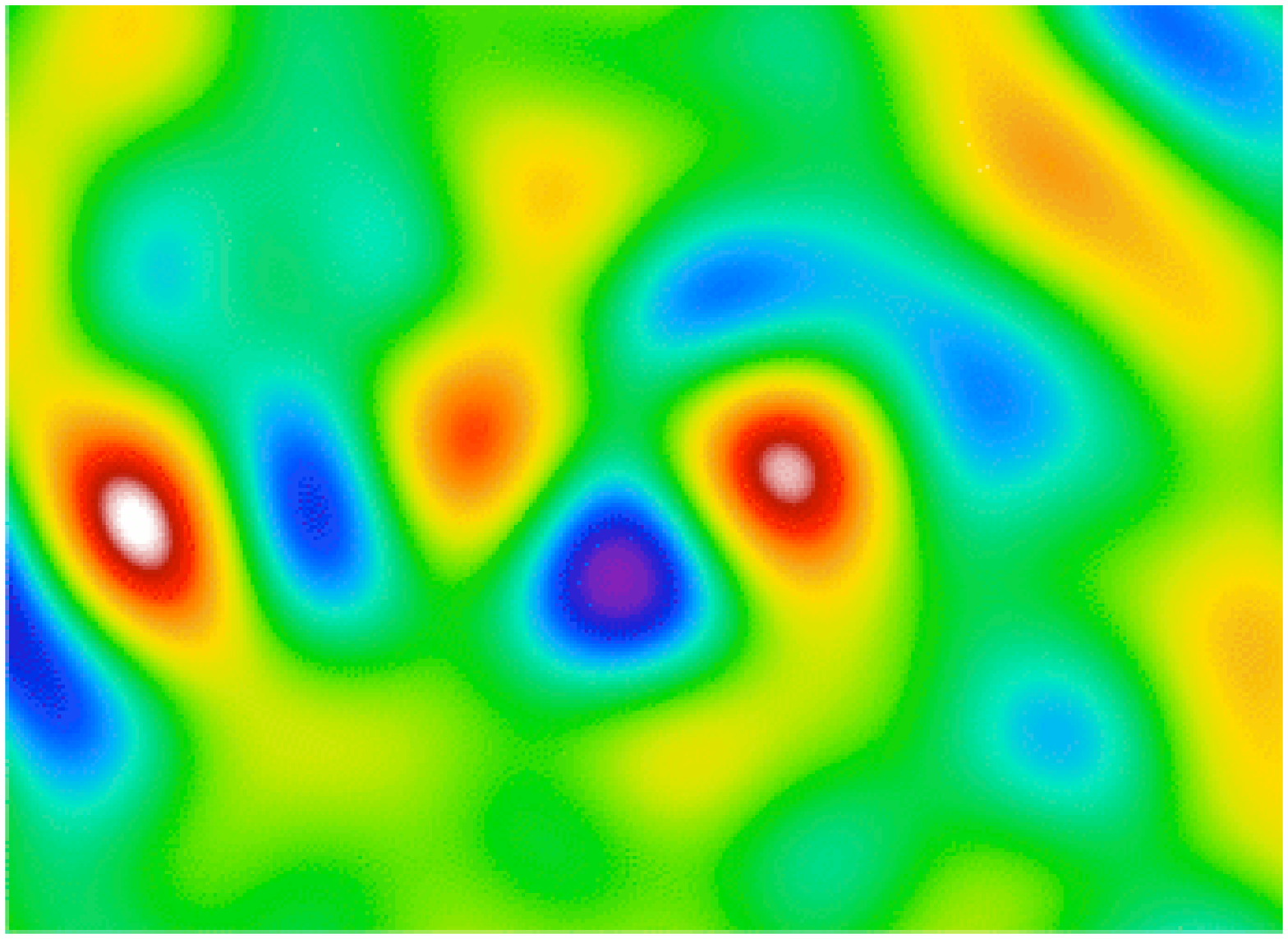} & $k<50$ &  \includegraphics[angle=270,width=6cm]{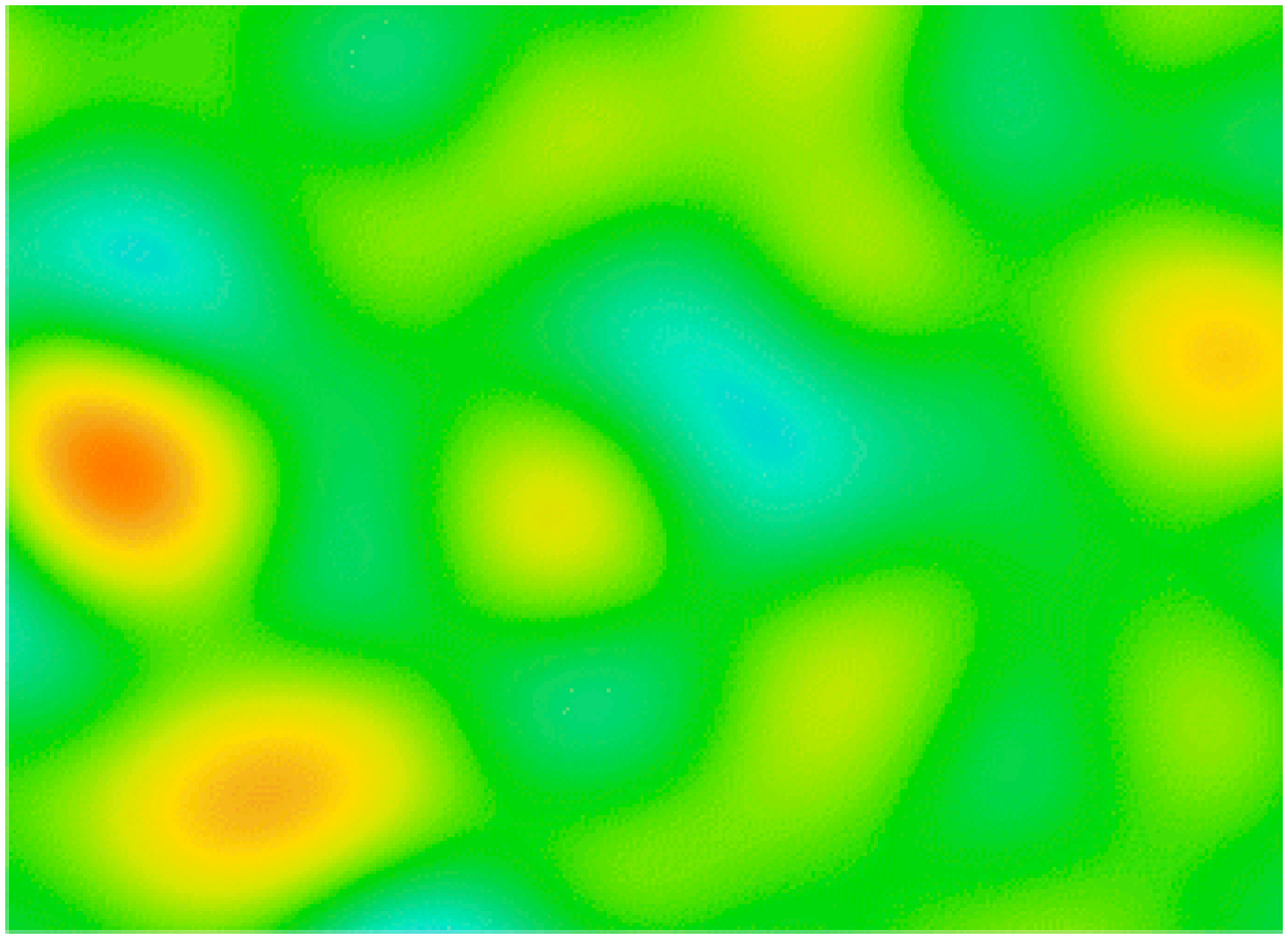} 
\tabularnewline
\includegraphics[angle=270,width=6cm]{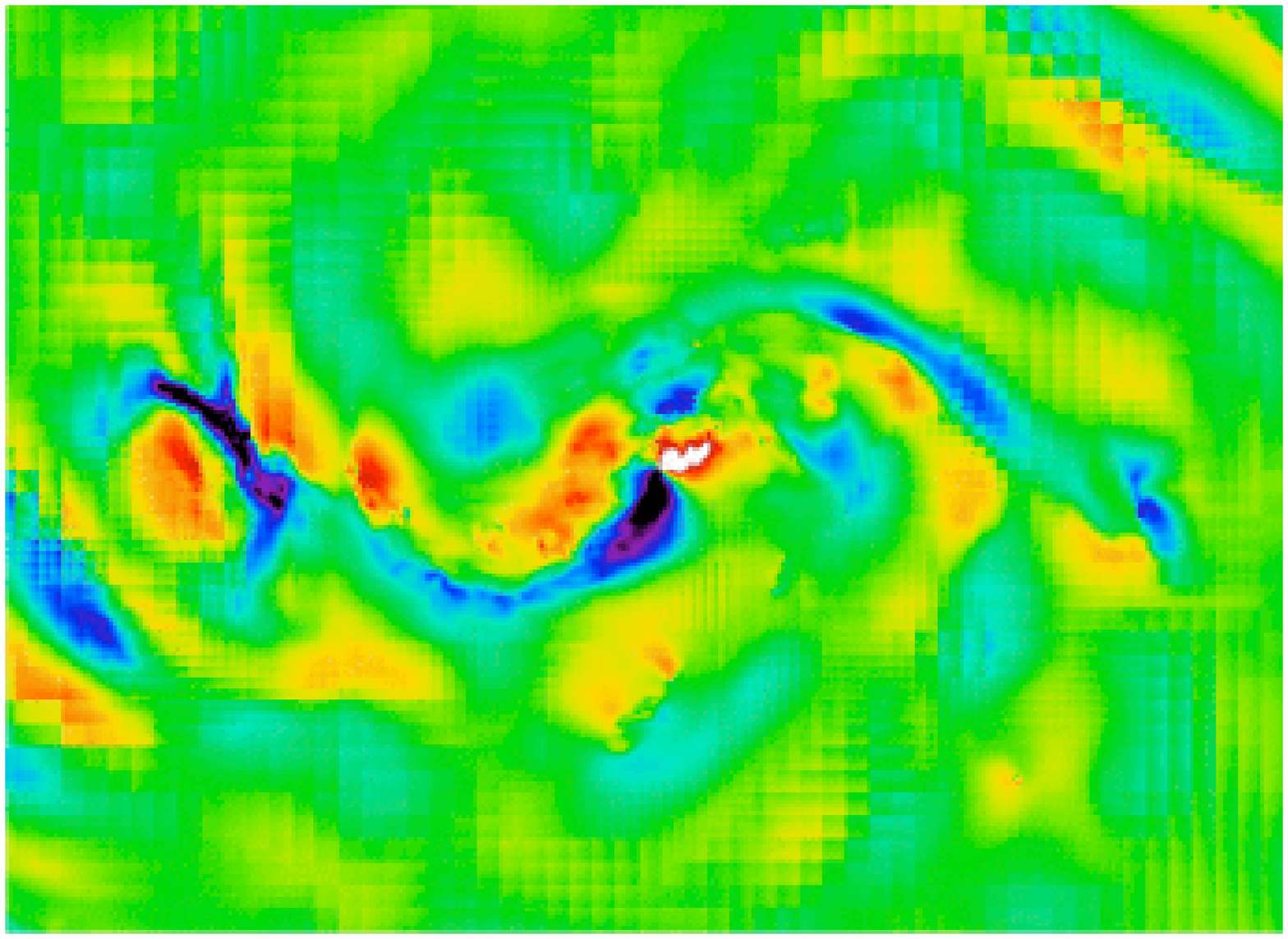}\vspace{.8cm} & $k>200$\vspace{.8cm} & \includegraphics[angle=270,width=6cm]{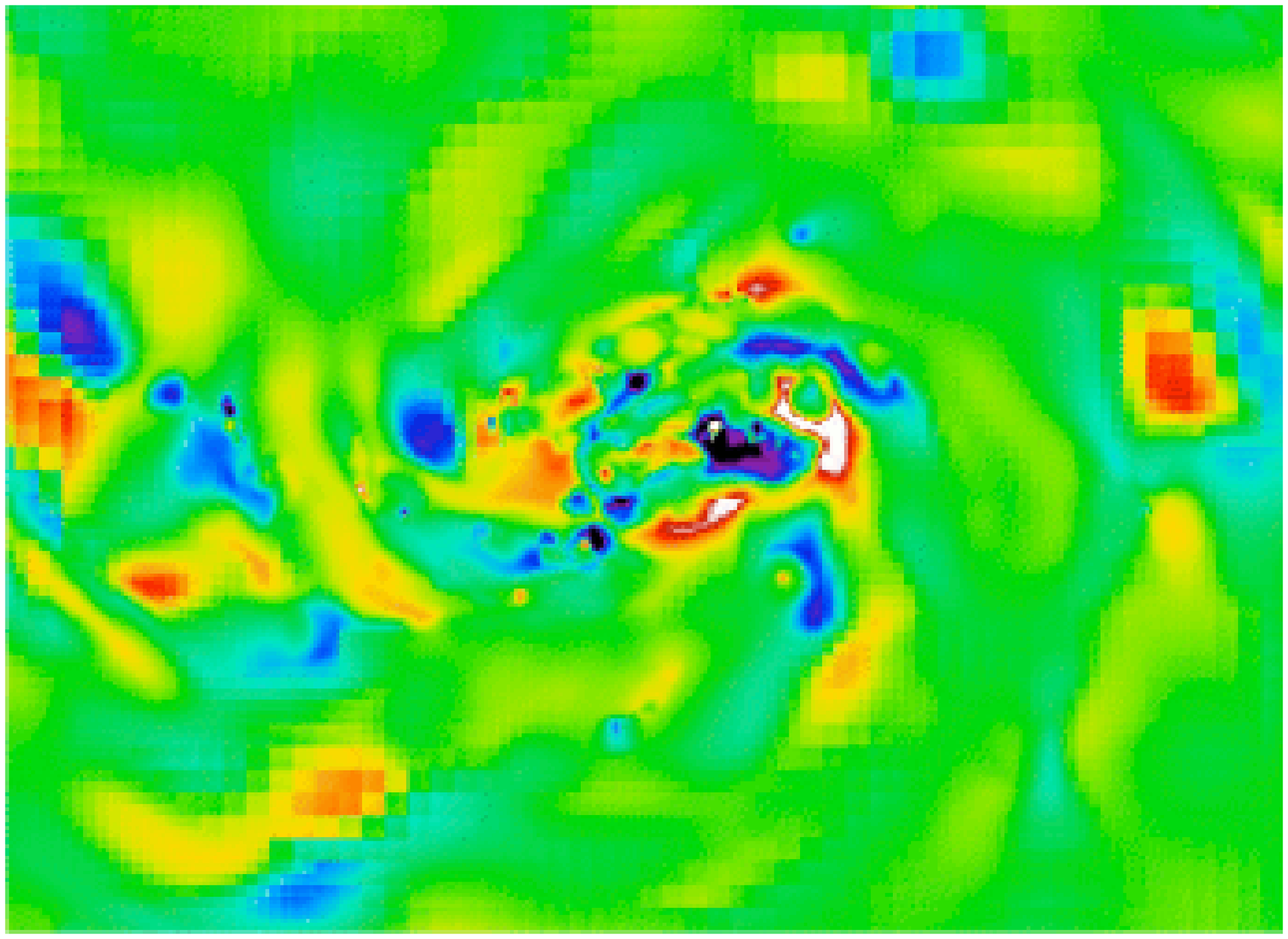} 
\vspace{.8	cm}
\tabularnewline
\multicolumn{3}{c}{
\includegraphics[width=12cm]{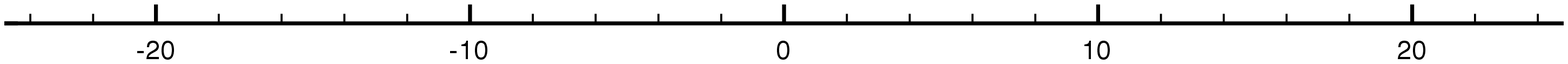}
}\vspace{-0.9cm}
\tabularnewline
\multicolumn{3}{c}{
\includegraphics[width=12cm]{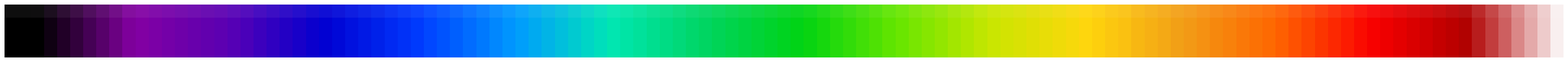}
}
\vspace{.45cm}\tabularnewline
\multicolumn{3}{c}{
Velocities (km~s$^{-1}$)}
\end{tabular}
\caption{Gas velocity fields at $t=254$~Myr. The in-plane $V_r$ component is shown in the left column, and the perpendicular component $V_z$ us shown to the right. Both are mass-weighted averaged along the line of sight. We show the total velocities (top), and the large-scale ($k<50$, middle) and small-scale ($k>200$, bottom) components, after zeroing the other wavenumbers in a Fourier decomposition. The break in the density power spectrum is at $k \sim 100$ and separates the two components shown separately here. The same colorbar is used for all maps. The field correspond to the face-on gas density image shown on Figure~\ref{fig:sim-density} in $7\times 4$~kpc snapshots (LMC-sized model with stellar feedback).}
\label{fig:sim-velocities}
\end{figure*}

\begin{figure}
\centering
\includegraphics[angle=270,width=8cm]{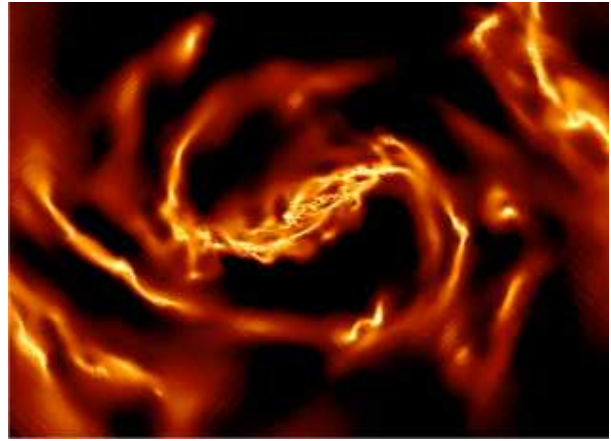}
\caption{Face-on gas density snapshot at $t=268$~Myr, with a $7\times 4$~kpc field of view, in the LMC-sized model with stellar feedback.}
\label{fig:sim2-density}
\end{figure}

\begin{figure*}
\centering
\begin{tabular}{>{\centering}m{6.1cm} >{\centering}m{1.25cm} >{\centering}m{6.1cm}}
$V_r$ & & $V_z$
\tabularnewline
\includegraphics[angle=270,width=6cm]{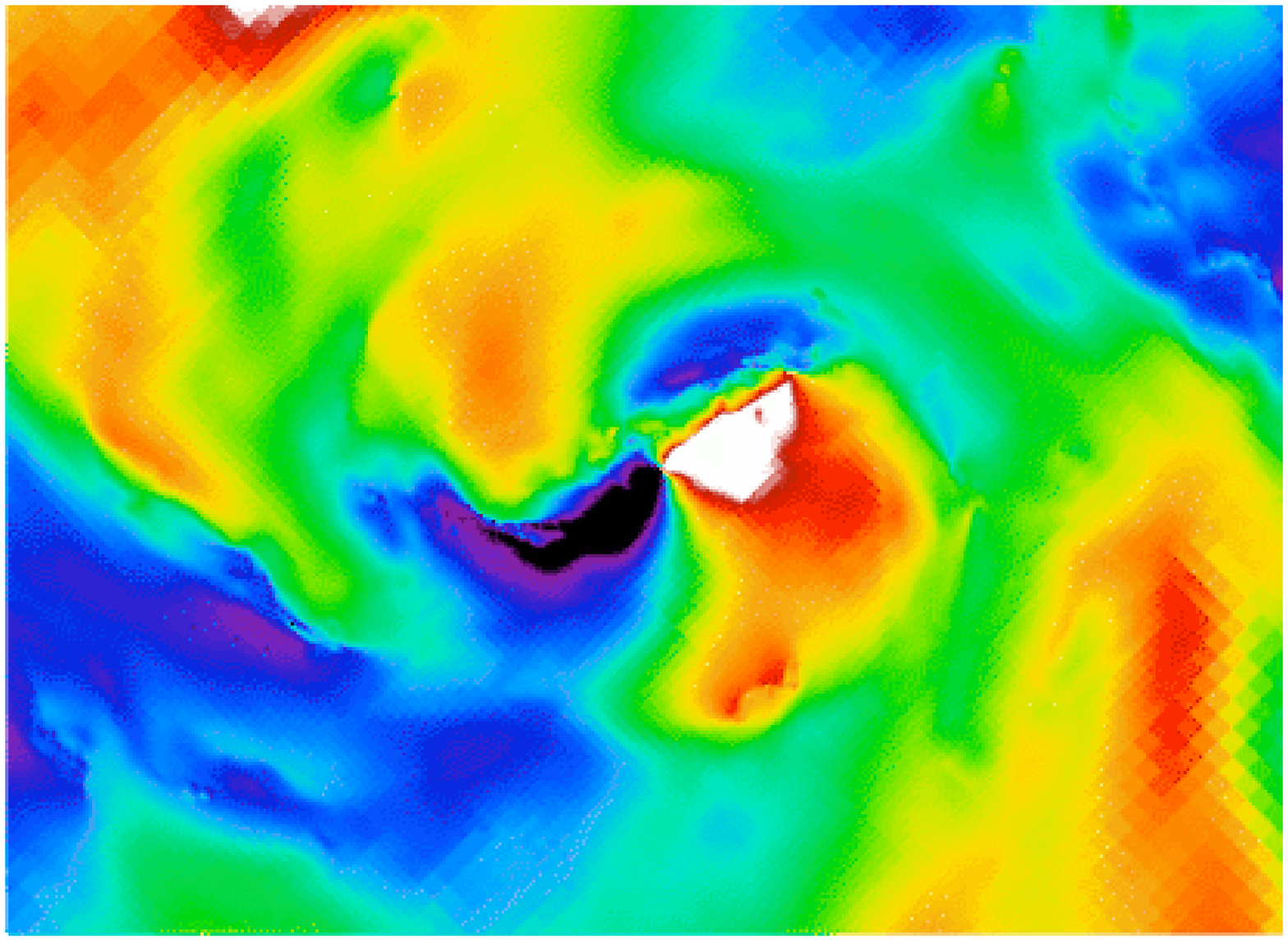} & total & \includegraphics[angle=270,width=6cm]{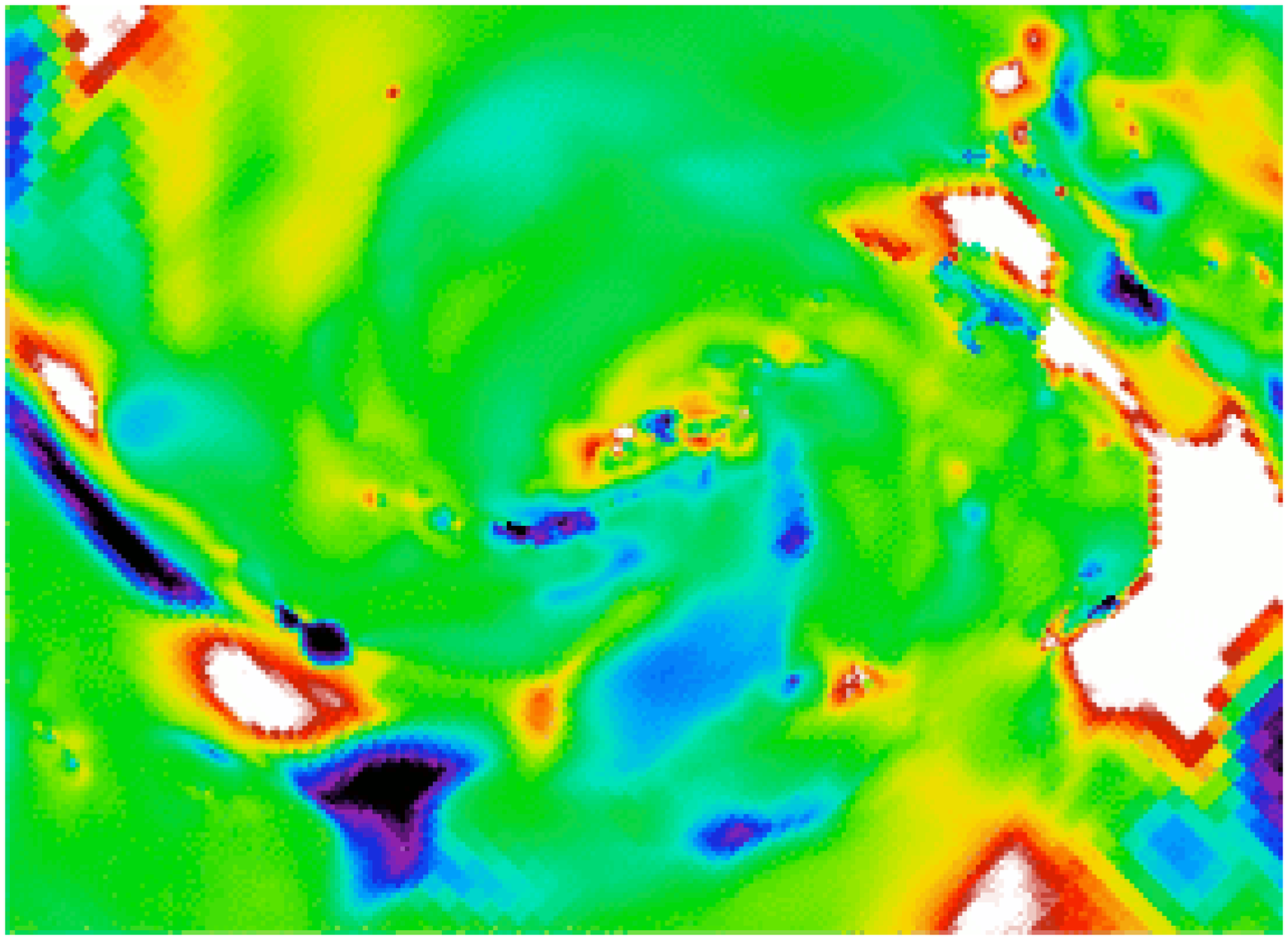} 
\tabularnewline
\includegraphics[angle=270,width=6cm]{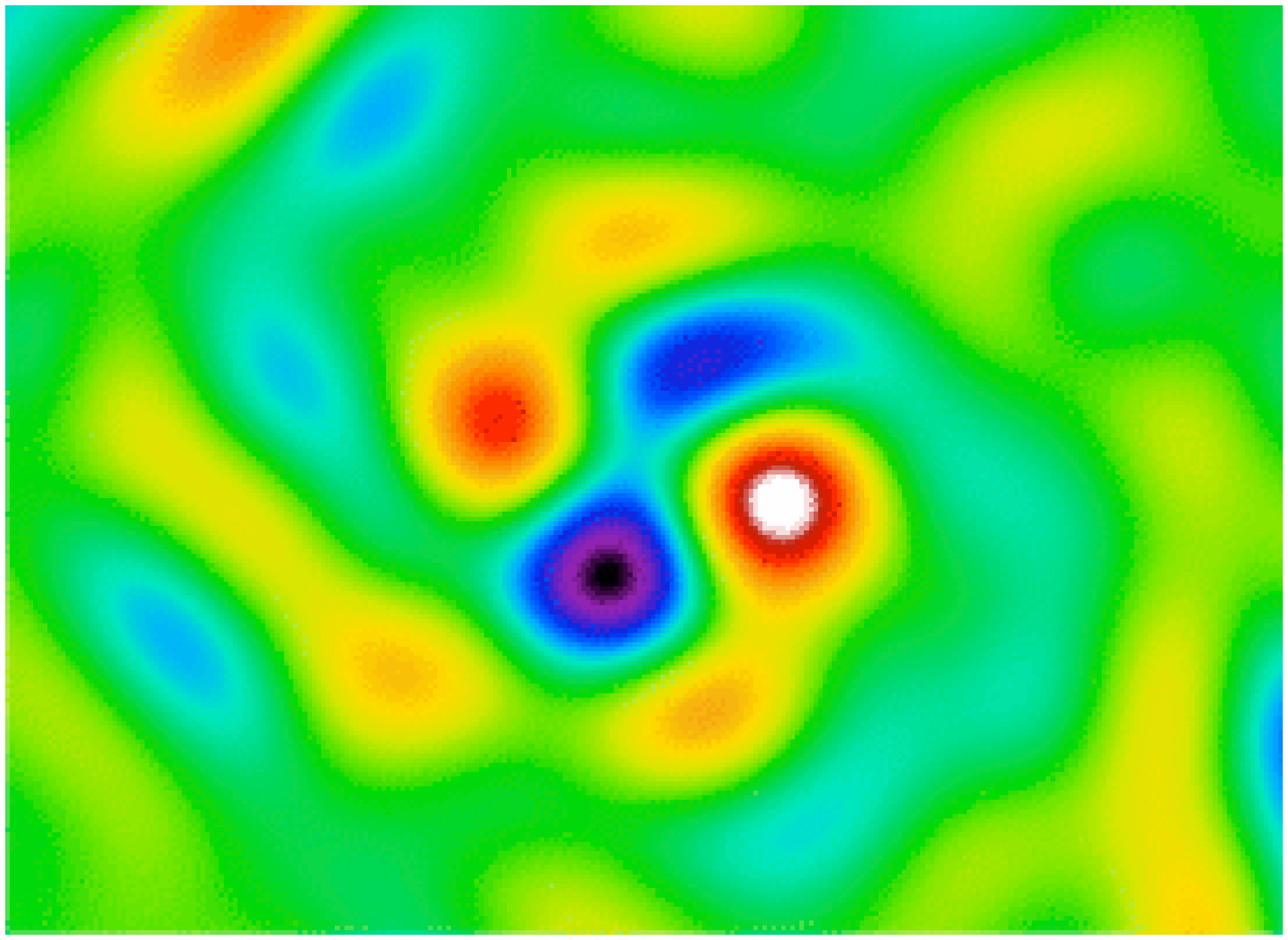} & $k<50$ &  \includegraphics[angle=270,width=6cm]{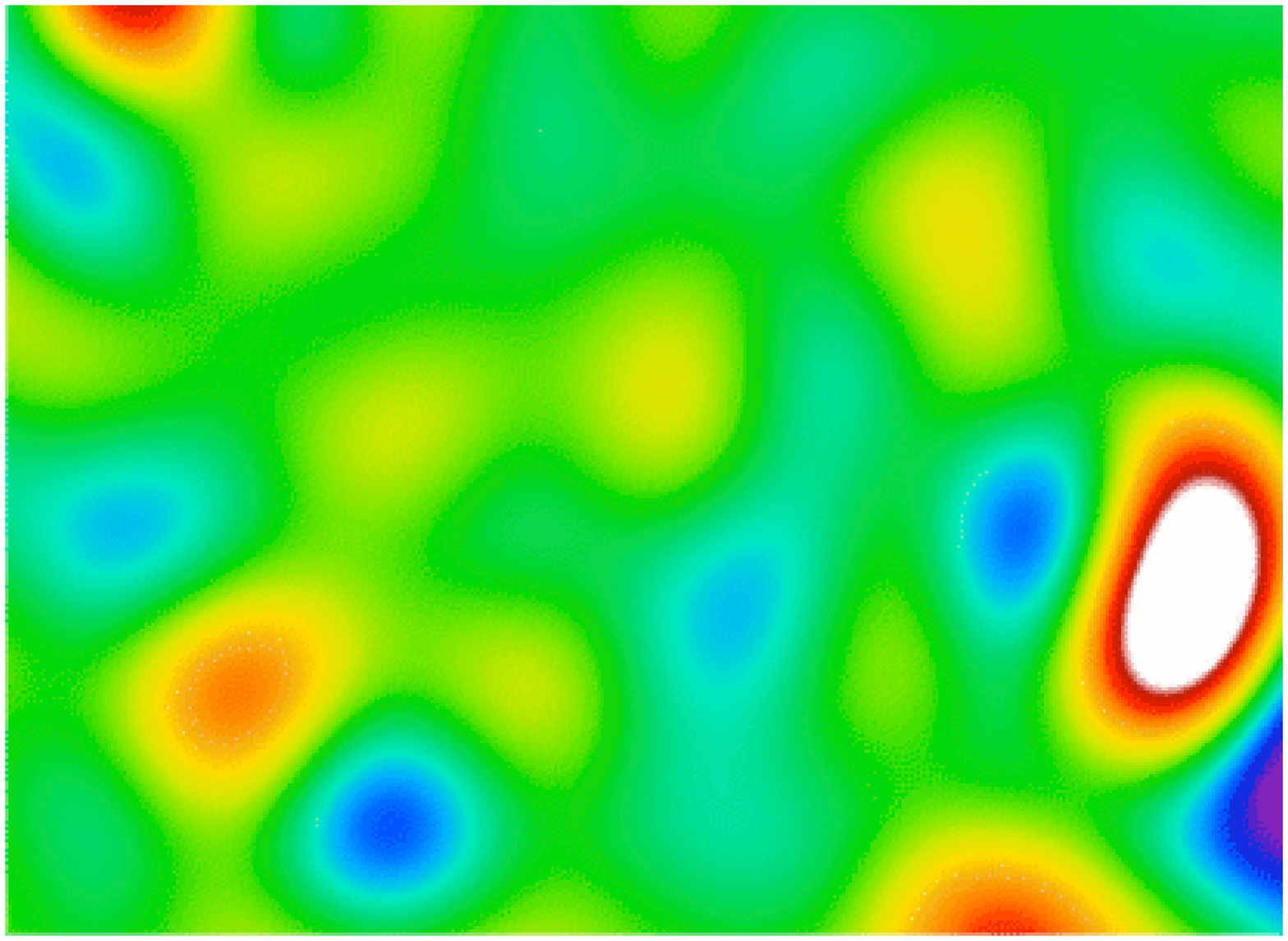} 
\tabularnewline
\includegraphics[angle=270,width=6cm]{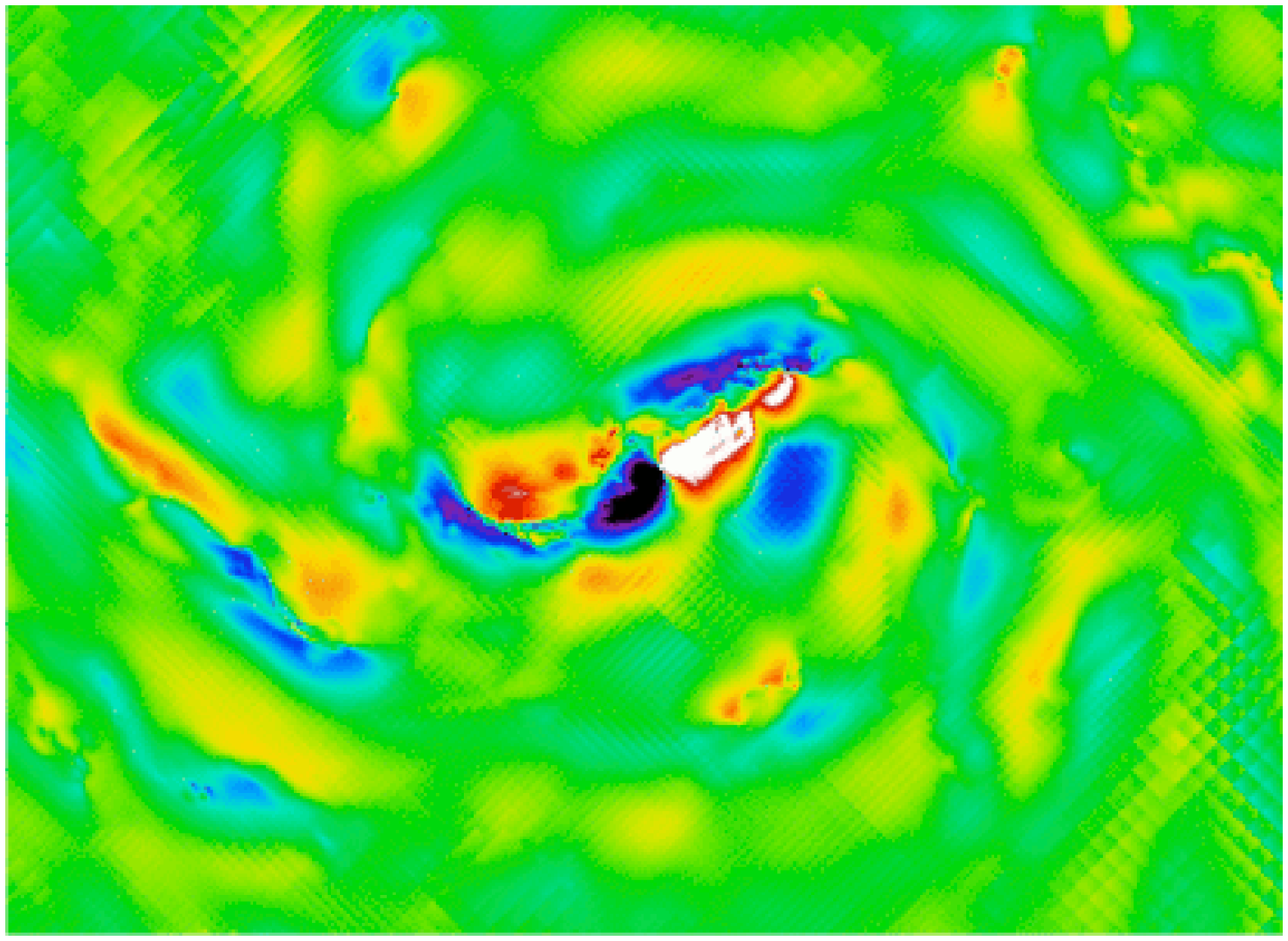}\vspace{.8cm} & $k>200$\vspace{.8cm} & \includegraphics[angle=270,width=6cm]{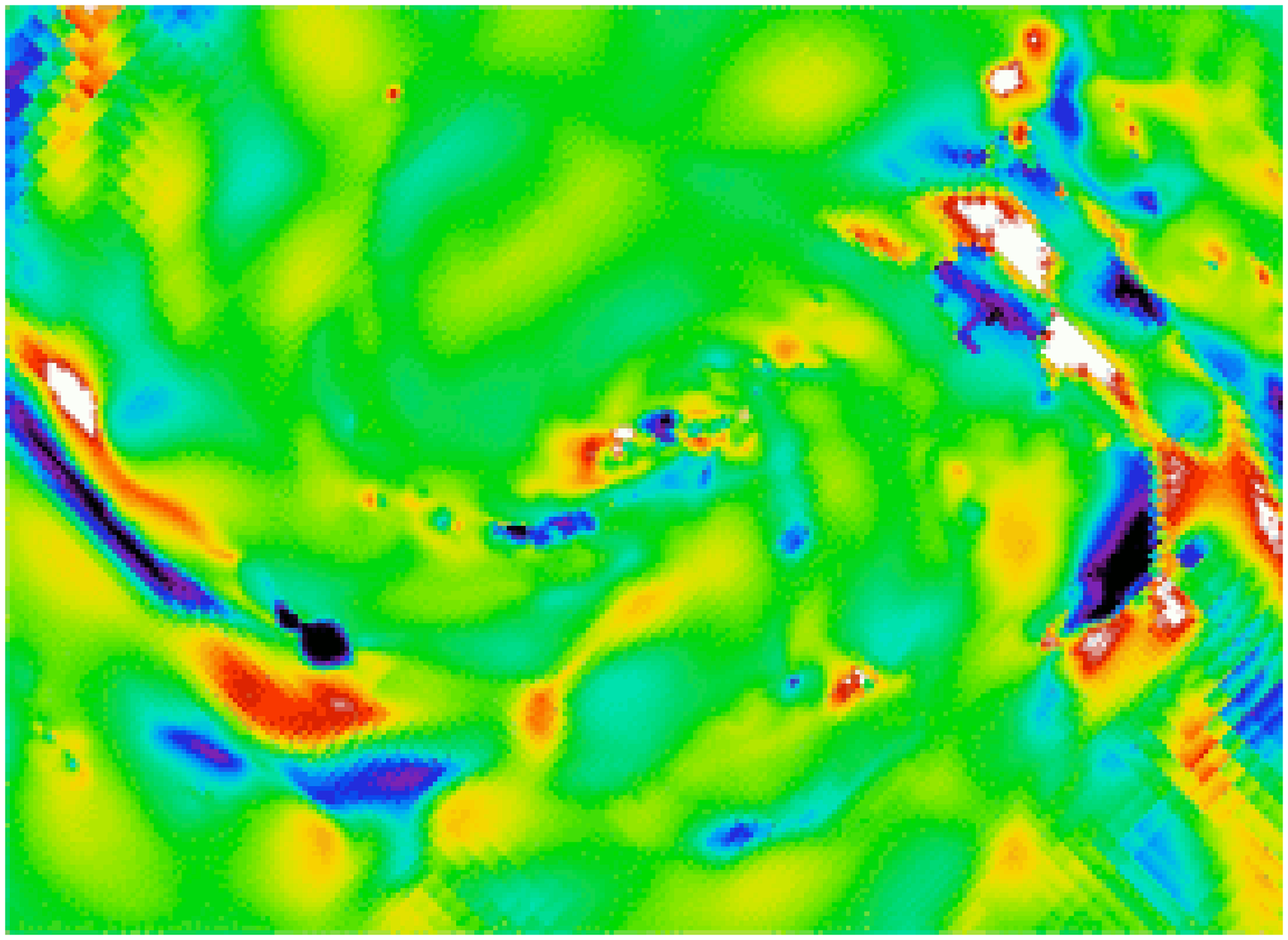} 
\vspace{.8	cm}
\tabularnewline
\multicolumn{3}{c}{
\includegraphics[width=12cm]{scale.eps}
}\vspace{-0.9cm}
\tabularnewline
\multicolumn{3}{c}{
\includegraphics[width=12cm]{color.eps}
}
\vspace{.45cm}\tabularnewline
\multicolumn{3}{c}{
Velocities (km~s$^{-1}$)}
\end{tabular}
\caption{Same as Figure~\ref{fig:sim-velocities} but at $t=268$~Myr, corresponding to the Figure~\ref{fig:sim2-density} density snapshot.}
\label{fig:sim2-velocity}
\end{figure*}

\subsubsection{Velocity fields and power spectra}
Maps of the in-plane radial velocity component $V_r$ and perpendicular component $V_z$ are shown on Figure~\ref{fig:sim-velocities} for the same snapshot as Figures~\ref{fig:sim-density} and \ref{fig:sim-density-zoom}, all at $t=254$~Myr. Another instant ($t=268$~Myr) is shown on Figure~\ref{fig:sim2-density} for the gas density and Figure~\ref{fig:sim2-velocity} for the velocity maps. The velocity maps are shown for the total velocity components, and for the low wavenumbers $k<50$ ($\lambda>300$~pc) and the high wavenumbers $k>200$ ($\lambda<70$~pc) separately. This separates the two regimes identified on the density power spectrum, where the transition occurs at $k \simeq 100$ or $\lambda \simeq 150$~pc.  These maps were built by zeroing the low-$k$ or high-$k$ components in the result of a Fourier transform of the velocity field, and recovering the velocities through an inverse Fourier transform. 

\medskip

The power spectrum of the three velocity components $V_r$, $V_{\theta}$ and $V_z$ is shown on Figure~\ref{fig:sim-velspec}. A single power law is found for the in-plane components. The perpendicular velocity $V_z$ follows the same power law on small scales, but its power spectrum flattens on large scales, with a transition at about the same scale length ($\sim$150~pc) as in the density structure spectrum. 

These differences in velocity power spectra correspond to higher power for the long-wavelength in-plane components than the long-wavelength perpendicular components of velocity. There are large in-plane motions at long wavelengths from spiral and bar-driven gas flows and relatively little perpendicular motions at long wavelengths to accompany them.

\subsubsection{A 3D cascade of turbulence on small scales}

On scales smaller than the gas disk scale height (100-200~pc), the ISM density has the density power spectrum expected for a 3D cascade of turbulence. The three-dimensional nature of the associated motions is confirmed by the high-wavenumber velocity maps, which show about the same amplitude for the in-plane motions and the perpendicular one. Typical average values of $<V_z/V_r>$ on various scales are given in Table~\ref{tab-vz-vr}. The small-scale motions are globally isotropic\footnote{The slightly higher values for the vertical velocity component correspond to low-density gas flows ejected above the disk plane by supernovae winds; the simulation without feedback has values of $<V_z>$ even closer to $<V_r>$ for high wavenumbers.}$^,$\footnote{Note that $V_r$ is slightly larger than $V_{\theta}$, so the in-plane motions are not perfectly isotropic. This is expected for instance if the gas tends to follow epicyclic motions, for which the ratio of the radial to tangential velocity dispersions is $2\Omega/\kappa$.}.

These results are overall consistent with a 3D cascade regime on the small-scale part of the density spectrum, initiated at scales around 150~pc, which is the gas disk scale height and the average Jeans length. This is naturally the case if the most gravitationally unstable scale, namely the Jeans length, corresponds to the disk thickness. Indeed, the gas disk scale height is expected to be set by the gaseous Jeans length in nearby disk galaxies \citep[]{E01,dutta}; this is also the case in high-redshift disk galaxies, which are more turbulent, have larger Jeans lengths, and have thicker gas disks at the same time \citep{EE06,genzel06,FS06,bournaud08b}. We checked this in our simulations by computing a mass-weighted average of the velocity dispersion $\sigma_{av}$ (8 km~s$^{-1}$), the average surface density $\Sigma_{av}$, and then an ``effective'' Jeans length $1/k_{\mathrm{Jeans,eff}}=\frac{\sigma_{av}^2}{\pi G \Sigma_{av}} = 213$~pc (locally, the Jeans length can significantly differ from this average value, as the density and velocity dispersion vary). This is consistent with the disk scale height being set by the Jeans length.

\begin{table}
\centering
\caption{\label{tab-vz-vr} Mass-weighted average values of $<V_z/V_r>$ for the large scale, low scale, and total velocities (LMC-sized model with feedback).}
\begin{tabular}{lcc}
Time & 254~Myr & 268~Myr 
\smallskip\\
small scales ($k>200$) & 1.13 & 1.07 \\
large scales ($k<50$) & 0.16 & 0.14 \\
total velocities (all $k$) & 0.92 & 0.89 \\
\end{tabular}
\bigskip
\end{table}

\subsubsection{A 2D inverse cascade on large scales?}

\begin{figure}
\centering
\includegraphics[angle=270,width=8cm]{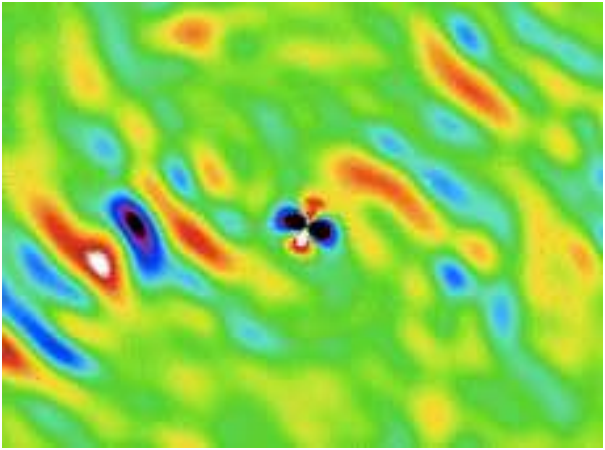}
\caption{Vorticity map for the large-scale ($k<50$) components, at $t=254$~Myr (LMC-sized model with stellar feedback). The scale ranges from $-10^{10}$~pc$^2$~Myr$^{-1}$ (black) to $10^{10}$~pc$^2$~Myr$^{-1}$ (white) with the same color bar as velocity fields.}
\label{fig:vorticity}
\end{figure}

\begin{figure}
\centering
\includegraphics[width=7cm]{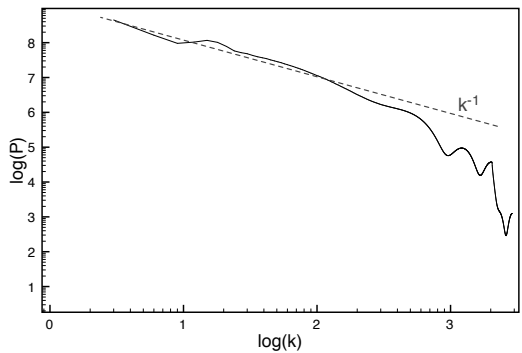}
\caption{Enstrophy spectrum at $t=254$~Myr for the LMC-sized model with stellar feedback. The wavenumbers are divided by a factor of 2 compared to the density and velocity power spectrum (Fig.~\ref{fig:sim-powspec} and \ref{fig:sim-velspec}). Scales larger than 100-200~pc (here $k \simeq 2$) have a -1 power law spectrum: the amplitude of vortices of various sizes is just as expected for two-dimensional turbulence.}
\label{fig:enstrophy}
\end{figure}

The low-wavenumber (large-scale) structures have a flatter density power spectrum. They also have different kinematic properties. The large-scale component of the velocity fields is dominated by in-plane motions. Perpendicular motions are much weaker. The low-wavenumber component of $V_z$ shows only modest peaks in Figure~\ref{fig:sim-velocities}, and these peaks are generally found in low-density regions (they generally correspond to gaseous fountains above the disk plan, rather than vertical motions in the mid-plane).

The motions corresponding to the large-scale regime of the density power spectrum are quasi-2D, highly anisotropic (Table~\ref{tab-vz-vr}). This is not because the disk rotation dominates these motions: this applies to $V_r$ as much as $V_{\theta}$, and the $<V_z/V_r>$ ratio is almost unchanged when we remove the lowest wavenumbers $k\leq3$ tracing global rotation. The power spectra of the three velocity components (Fig.~\ref{fig:sim-velspec}) actually shows that the perpendicular velocity is much lower than the in-plane components for all wavelengths larger than the disk scale-height.

The $\sim -2$ slope of the large-scale branch of the density power spectrum and the quasi-2D nature of the associated motions suggest that we are observing something like a 2D inverse cascade of turbulence \citep{2Dturb}. We nevertheless have to explore the nature of these motions more accurately, to ensure that these properties are not a conspiracy from density waves and associated gas flows, which would then not be turbulent motions. 

The large-scale components of the in-plane motions (Figs~\ref{fig:sim-velocities} and \ref{fig:sim2-velocity}) show a central $m=2$ mode, which is the characteristic response to a bar+arms system of density waves \citep[e.g.,][]{combes, athanassoula}. This response is however quite asymmetric, and does not dominate the large-scale motions outside of the central kpc. Overall, outside the central kpc, chaotic motions dominate the large-scale velocity components (of course superimposed on the disk rotation pattern). In particular the comparison of the gas density and velocity maps shows large vortices around dense gas clumps: for instance, the densest clump seen to the left of Figure~\ref{fig:sim-density} is clearly associated with an in-plane eddy, with positive and negative peaks of $V_r$ surrounding the position of this dense clump. Such signatures are reminiscent of a Rosby Wave instability \citep{lovelace99, varniere-tagger06}, which is observed in pure 2D disks but also in thick disks or 3D systems \citep{meheut2010} and could thus take place in our quasi-2D system. Such motions would be quasi-2D turbulence, which is more than a large-scale flow in a density wave, but which may be induced by a density wave of other large-scale forcings.

The vorticity map of the large-scale velocity component ($k<50$), after subtraction of the $m=0$ rotation pattern, is shown on Figure~\ref{fig:vorticity}. It shows a number of vortices with various sizes and strengths throughout the disk. The power spectrum of the enstrophy is shown in Figure~\ref{fig:enstrophy}. Enstrophy is the integral of the square of the vorticity: $\mathcal{E} = \int _\mathcal{S} || \vec \nabla \times \vec v ||^2$. The enstrophy has a power-law power spectrum throughout the entire range of 2D scales, like the in-plane motions shown before. The slope of this power spectrum is -1, just as expected for 2D turbulence, as found in two-dimensional experiments \citep[e.g.,][]{petersen06,petersen07}. This enstrophy spectrum means that the quasi-2D part of the density power spectrum is made-up of vortices with size and strength distributed as expected for 2D-turbulence: this is unlikely to simply result from non-turbulent flows along density waves.

\medskip

The properties of the gaseous motions on scales larger than the disk scale height thus contain a component from turbulence in a quasi-2D medium. The origin of this turbulence could be a combination of long-range forcing from spiral waves and other global disturbances, and an inverse cascade from smaller scales, where energy is put into the ISM by gravitational instabilities on the Jeans length and stellar feedback.  Energy injection at the Jeans length is consistent with our observation of vorticity around the gravitating clumps and with the quasi-2D power spectrum on scales larger than the Jeans length. \citet{wada02} already mentioned that the properties of turbulent motions in their galaxy models were consistent with an inverse cascade of turbulence. Their models were purely two-dimensional, while we here suggest a similar inverse cascade in a three-dimensional gas disk, which is thin but has a well resolved scale height. 

The large-scale motions probably cannot simply result from a single inverse cascade with energy input only at the Jeans scale. Indeed, the growth rate of gravitational perturbations in a disk decreases only as $1/\sqrt{\lambda}$ where $\lambda$ is the scale of the perturbations, so that long-range forcing at scales larger than the typical Jeans scale length remains significant. At least, the gravitational coupling of stellar and gaseous components and the bi-component stability \citep{jog} should result in a large range of unstable scales, ranging typically from the gaseous Jeans scale to the stellar Jeans scale \citep{E95}. There should then be a range of unstable scales injecting energy into the inverse turbulence cascade. While the expected properties of 2D turbulence, such as the enstrophy spectrum, are recovered in our simulations, one can note that the slope of the density power spectrum in this regime is slightly higher than expected from pure two-dimensional turbulence (a $-2$ power law, instead of the $-5/3$ exponent expected for purely two-dimensional and uncompressible turbulence). The non-zero thickness of the gas disk and the relative importance of long-range forcing in a gaseous+stellar disk might cause this.

The role of long-range gravitational forcing and the coupling of the gaseous and stellar components also appear by comparing our simulations with the AMR models in \citet{tasker09}. The stellar disk is modeled with a rigid analytical potential in their simulations. Gas clumps form everywhere in their disk with a relatively peaked size distribution and a constant separation of about one Jeans scale. The Jeans length is also a characteristic scale in our model: it sets the disk scale height and gas clouds along spiral arms are also often separated by $L_J$ (Fig.~\ref{fig:sim-density-zoom}), but gaseous structures overall form over a wider range of sizes in our models. The inclusion of both stellar and gaseous gravity on large scales appears crucial to realistically reproduce the full spatial range of the main ISM structures from long spiral arms and large vortices to smaller clumps and shocks, even if ISM structures on small scales can be realistically reproduced without stellar gravity \citep[e.g.,][]{avillez07}.

\bigskip

We have analyzed above the LMC-sized simulation with star formation and stellar feedback. In this simulation, the properties of ISM turbulence seem mainly driven by gaseous and stellar gravity. However, the comparison to the model without stellar feedback below will highlight a fundamental role of feedback in maintaining a steady state in the density distribution of the ISM. 

\subsection{Role of star formation feedback in gas disk properties}

\begin{figure}
\centering
\includegraphics[angle=270,width=8cm]{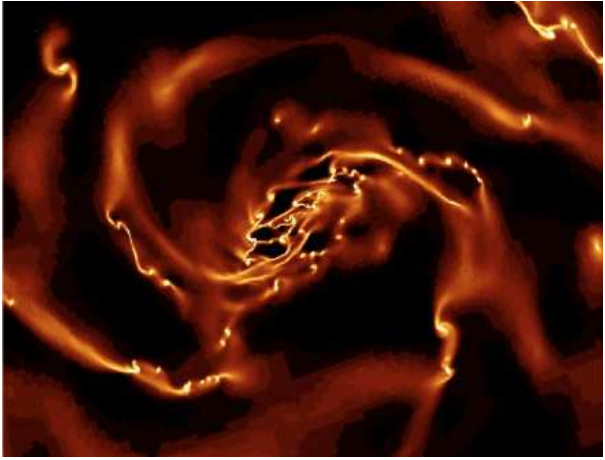}
\caption{``Gravity-only'' LMC-sized model without star formation and feedback: face-on view of the gas density at $t=343$~Myr.}
\label{fig:gravity-map}
\end{figure}

\begin{figure}
\centering\vspace{.4cm}
\includegraphics[width=8cm]{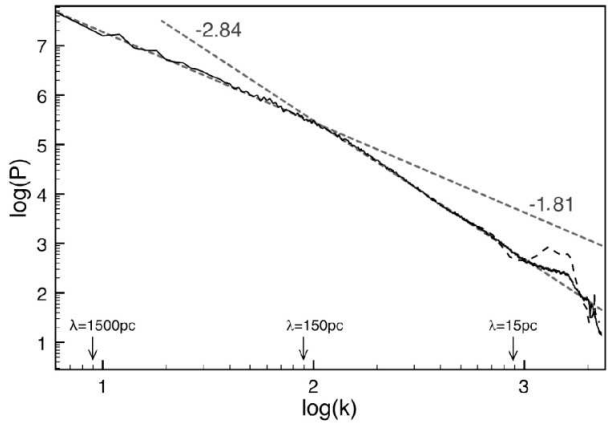}
\caption{Face-on gas density power spectrum of the ``gravity-only'' model without stellar feedback at $t=258$ (solid lines) and $343$~Myr (dashed, shown at small wavelengths only for clarity).
}
\label{fig:gravity-spec}
\end{figure}

The ``gravity-only'' simulation (without star formation and feedback) shows a relatively similar distribution of gaseous structures, and a relatively similar density power spectrum compared to the model with feedback (see Figures~\ref{fig:gravity-map} and \ref{fig:gravity-spec}), at least at $t=238$~Myr and at scales larger than 5-10~pc. Besides suggesting that our earlier results do not crucially depend on a specific feedback scheme, it also indicates that the ISM structuring and the pumping of turbulent energy into the turbulent cascades results mainly from gravitational processes and is not primarily driven by supernovae explosions. The disk scale-height and the disk substructures are unchanged in this gravity-only model.

\medskip

Feedback processes nevertheless play a fundamental role in maintaining this gravity-driven turbulence distribution. We can indeed note already at $t=238$~Myr in Figure~\ref{fig:gravity-spec} that the ISM power spectrum shows an excess of very small-scale structures (below 5-10~pc) and the gas density map shows the associated small and dense gas clumps. Over longer timescales the power spectrum becomes less realistic as the large scales are gradually emptied and the bump at very small scale increases: gas dissipates its energy and accumulates in tiny dense bullets from where it is never removed (Fig.~\ref{fig:gravity-map}). 

The role of feedback appears when comparing the density power spectra in our two models: feedback disrupts the dense smallest-scale structures, and supernovae bubbles expand up to the disk scale height. Density maps show supernovae-driven bubbles, all of which are smaller than or comparable to the disk scale-height. But this does not mean that the size of supernovae bubbles regulates the scale height: the measured scale-height and the break in the density power spectrum are not changed by feedback; the scale height is mostly regulated by gravitational processes, and is about the Jeans length. But feedback disrupts structures on the smallest scales and expands them up to the Jeans length, where gravitational processes can initiate a new cycle of 3D turbulent cascade towards small scales and 2D inverse cascade towards large scales.

Overall, the gas disk structure and its scale height are largely regulated by gravitational instabilities, but gas structures and energy dissipation on the smallest scales are also regulated by star formation feedback. The break in the face-on density power spectrum and the scale height measured on edge-on projections both correspond to the average Jeans length. 2D and 3D motions are initiated by instabilities around this scale length and large-scale forcings. Feedback processes prevent the otherwise inevitable accumulation of gas in tiny and dense bullets, disrupt small-scale structures and refill the turbulent cascades initiated at the Jeans length. This cycle maintains the ISM density distribution in a steady state. 

Observations have already suggested that turbulent motions are primarily driven by gravitational instabilities such as spiral arms and molecular clumps \citep{E03}. Based on the analysis of the density distribution, these authors pointed out the role of star formation feedback in breaking apart small dense structures to preserve a steady state. \citet{agertz09} also proposed from simulations that the main driver of ISM turbulence could be gravitational forcing, with nevertheless a significant contribution from star formation (see also Dib, Bell \& Burkert 2006).

\medskip

Gravity-only models without star formation and feedback initially produce a realistic density distribution in the ISM over a couple of disk rotation periods. But on longer timescales and over large numbers of disk rotations, the inclusion of feedback in numerical models appears necessary to maintain a realistic density distribution. However, this applies only to high-resolution simulations where gas cooling is modeled down to $\sim 100$~K. Lower-resolution SPH simulations, and more generally models where the ISM is only modeled as a smooth stable gas at $\sim$$10^4$~K or more (e.g., the stabilizing EoS by Springel \& Hernquist 2003 and Robertson et al. 2004), would not be affected by catastrophic gas dissipation if they do not include feedback sources, as they do not resolve the turbulent and unstable nature of the star-forming ISM.

\subsection{Extension to massive and gas-rich galaxies}

The results described so far were for a low-mass disk galaxy, scaled to the LMC properties. To explore whether the highlighted properties can extend to more massive disk galaxies and/or disk galaxies with different gas fractions, we have applied the same analysis to a high-redshift galaxy model described in \citet{bournaud10}. This is a massive galaxy (baryonic half-mass radius of 10~kpc and circular velocity of 280~km~s$^{-1}$) with a 50\% gas fraction, modeled with a 2~pc resolution. As typical star-forming galaxies at $z \sim 2$, it has a high gas fraction, a very unstable disk with high turbulent speeds (a few tens of km~s$^{-1}$) and giant complexes of star formation in a thick ($h_{z} \sim 1$~kpc) gas disk (see observations by \citet{EE06,E07,genzel06} and simulations by \citet{semelin,BEE07,agertz10,ceverino10}). 

\begin{figure}
\centering
\includegraphics[width=8cm]{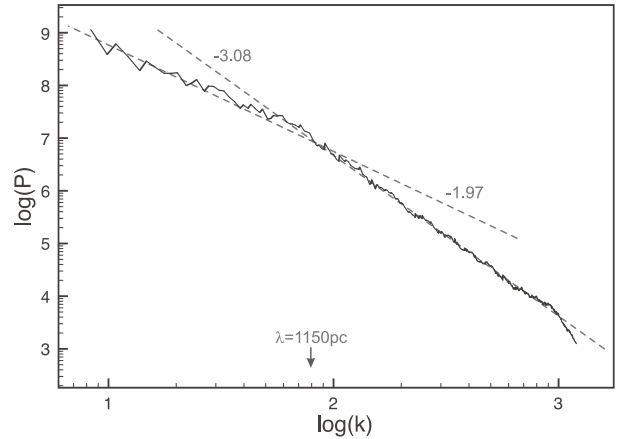}
\caption{Gas surface density power spectrum in a massive gas-rich galaxy model.}
\label{fig:highz}
\end{figure}

The density power spectrum is shown on Figure~\ref{fig:highz}. It is consistent with a double power law, with the break occurring at wavelength around 1~kpc which is the disk scale height and the typical Jeans length in such galaxies. We measured a total ratio of velocity dispersions of $<V_z/V_r> = 0.88$, with $<V_z/V_r> = 0.26$ on large scales and $<V_z/V_r> = 1.07$ on small scales. The large-scale motions are still significantly non-isotropic, even if the low wavenumber velocities are somewhat more isotropic than in our LMC-sized model, which could be caused by the disk being thicker. The properties in the density distribution and chaotic motions found in our LMC-sized models thus seem to apply, at least qualitatively, to galaxies with different masses, rotation curves and gas fractions.

\subsection{Resolution requirements}\label{resol}

Our fiducial LMC-sized model with star formation and feedback has been re-done with $l_{max}=11$, i.e. a spatial resolution of 13~pc, which means that the disk scale height is resolved with $\sim$10 resolution elements, instead of 100-200 resolution elements at the maximal refinement level $l_{max}=15$. The density threshold for star formation was reduced by a factor of 80 to keep the resulting star formation rate similar. 

The global density distribution of this reduced-resolution model is not very different from the initial case, except that the highest density structures in the PDF are removed; otherwise the PDF is only moderately shifted towards lower densities and has about the same dispersion (Fig.~\ref{fig:sim-pdf}). However, the velocity structure is significantly different. The initial model had almost isotropic velocity dispersion $\sigma_z /\sigma_r \simeq 0.9$ (the smallest scales begin the most isotropic ones). Interestingly, the reduced-resolution model has much less isotropic motions, with $\sigma_z/\sigma_r =0.32$ (at $t=262$~Myr and over all wavenumbers; the ratio was $\sim$0.9 in the high-resolution run). We thus note that a very high resolution is required to resolve the isotropy or non-isotropy of turbulent motions in disk simulations, and that a few ($\sim$10) resolution elements per disk scale height are not sufficient. Our interpretation is as follows: vertical motions can be initiated or amplified when instabilities occur off the midplane at heights $z$$>$0 or $z$$<$0, when the Jeans length becomes locally smaller than the disk scale height (which happens in dense regions or low-velocity dispersion regions). This requires to resolve, locally, several Jeans length per scale height. Then, each Jeans length itself needs to be resolved by a least a few resolution elements: 4 resolution elements per thermal Jeans length is a strict minimum imposed in our model and our EoS generally results in 5-10 elements per thermal Jeans length, and the number of resolution per thermal+turbulent Jeans length would be even higher. Thus, capturing instabilities at $z > 0$ and the associated triggering of vertical motions can require a few tens of resolution elements across the gas disk scale height. When the disk thickness is marginally resolved, the non-isotropy of gas motions can be artificially increased at small scales. In this case, each gas clump is mostly spinning around its minor axis (see Agertz et al. 2009a) generally aligned with the spin axis of the entire galaxy (but see Tasker \& Tan 2009).

\section{Summary and conclusions}

\subsection{Properties of ISM turbulence, role of gravity and feedback}

This paper has presented an LMC-sized models for comparison with observations of the LMC (Block et al. 2010) and a high-redshift clumpy disk model. The LMC-sized model does not match all individual structures observed in the LMC -- this is not the initial purpose and the LMC has many low-density ionised filaments, and other substructures in low-density regions, that are out of reach of our mass resolution.

The model is successful in reproducing the double power law shape of the ISM density power spectrum observed in the LMC (Elmegreen et al. 2001 and Block et al. 2010) in a system with a similar mass, gas fraction and rotation curve. We analyzed the 3D velocity structure of the models, which could not be done in observations, to understand the origin of this density substructures.

Our results indicate that a dual turbulence cascade could take place in our simulations, and by extension in the real ISM. The scale height of the gas disk is set by the Jeans scale length, which is the most gravitationally unstable scale. Gravitational instabilities at/around this scale length inject energy in turbulent motions. These motions follow a classical 3D cascade of isotropic turbulence towards smaller scales, on which they form a hierarchy of gas clouds, GMCs and substructures therein. Turbulent motions on scales down to and inside GMCs thus appear to be from self-gravity. Nevertheless, stellar feedback plays a major role in disrupting dense structures on the smallest scales and in maintaining the turbulence cascade in a steady state.

On scales larger than the gas disk scale height, very anisotropic turbulent motions with a shallower density power spectrum are observed. Long-range gravitational forces are probably not negligible, but the vorticity and enstrophy properties on these large scales are as expected for (quasi-)2D turbulence. The large scale motions are not limited to global disk rotation and streaming flows along density waves. This inverse cascade appears to be initiated mainly by the same gravitational instabilities at the Jeans length as the direct 3D cascade, because it is present with and without stellar feedback.

While models without self-gravity do not find a characteristic scale for energy injection into ISM turbulence \citep{joung-ml}, we identify a major role of gravitational processes around the Jeans length. Large-scale forcing and stellar feedback at small scales are important processes in regulating the ISM properties too, but the Jeans length is the scale at which chaotic gas motions change from quasi-2D to 3D isotropic turbulence, and at which a break occurs in the density power spectrum. Log-normal gas density PDFs have been found in models of supernovae-driven ISM turbulence without self-gravity \citep[e.g.][]{avillez02}, but without having the observed power spectrum reproduced. The power spectrum analysis in our models and in LMC observations rather suggests that the energy injection in the ISM is largely from gravitational processes \citep[such as gravitational instabilities and inward mass accretion, ][]{EB10,KH10} with a regulating role of feedback on small scales.

\medskip

We have shown that similar properties for ISM substructures and turbulence, at least qualitatively, are found in models of galaxies with higher masses and higher gas fractions. This includes the extreme case of high-redshift ($z \sim 2$) galaxies with thick, highly turbulent and very clumpy gaseous disks. An interesting implication for observers is that the isotropy of the gas velocity dispersion is strongly dependent on the scale, and hence on the resolution of an observation.

\subsection{Modeling ISM substructures in galaxy simulations}

We have explored ISM modeling with a barotropic equation of state (EoS) corresponding to gas in thermal equilibrium between a standard radiation field with the main atomic and fine-structure cooling processes. We have shown that this EoS employed in an AMR hydrodynamic code with a quasi-Lagrangian refinement strategy reproduces a realistic power spectrum for the ISM density distribution. It does not form a truly multiphase medium, since the pressure is only a function of density. It nevertheless a clumpy ISM, with cold and dense clouds embedded in a warmer diffuse phase, and the resulting density distribution is consistent with that of the real, multiphase ISM.
This technique is efficient to reach high spatial and mass resolutions: the computation of cooling and heating rates is not numerically costly, but the absence of out-of-equilibrium, colder or warmer gas in this model prevents the timescale to become extremely short, which would typically happen in parsec-scale models with complete cooling calculations.

Hydrodynamic models with cooling calculations produce a log-normal PDF for the ISM \citep{wada07,tasker09,agertz09}, which is retrieved in our model over a larger dynamical range. The PDF alone does not guarantee that the correct mass fraction is distributed in structures of various sizes and densities. Our model with a mass and rotation curve similar to the LMC matches the density spectrum observed by Block et al. (2010) in the LMC. While all thermal processes are not explicitely computed, this at least means that the ISM modeled this way has realistic density distribution, with a correct number of gas structures of various sizes and a correct mass fraction therein, from the global galaxy-sized scales down to 5-10~pc, i.e. molecular clouds and their main substructures.

The ISM substructures are mostly powered by gravity-driven turbulence and instabilities around the Jeans scale length. Star formation feedback is however an essential ingredient, not only by structuring the ISM into shells locally, but also by providing an extra source of energy is needed to disrupt the smallest and densest structures that tend to be steadily filled by gas dissipation into compact clumps. Gravity, hydrodynamics, and a realistic thermal model are initially enough to produce a realistic structure distribution in the ISM: feedback seems relatively negligible over a couple of disk rotations, but becomes important over longer timescales to balance systematic gas dissipation. The source of energy that disrupt the structures at very small scales and refills the 3D cascade and the inverse 2D cascade at the Jeans scale is, in our model is supernova feedback. Other sources such as HII regions, stellar winds, and radiation pressure from young massive stars \citep{murray, kd10} could also contribute to maintain the ISM density distribution in a steady state. The global density structure of the ISM remains relatively unchanged in simulations where the disk scale height is marginally resolved, but modelling accurately the quasi-isotropic turbulent motions on small scales is found to require a very large number (at least a few tens) of resolution elements per disk scale height.

\section*{Acknowledgments}
This work was granted access to the HPC resources of CCRT and CINES under the allocation 2010-GEN2192 made by GENCI (Grand Equipement National de Calcul Intensif). We are grateful to John Scalo and Sebastien Fromang for suggestions, and to Oscar Agertz, Daniel Ceverino, Fran\c{c}oise Combes and Elizabeth Tasker for discussions and comments on the manuscript. DLB is indebted to Mr F.~Titi as well as the AVENG Group of Companies for their sponsorship. DLB thanks Roger Jardine, Kim Heller and the AVENG Board of Trustees. IP acknowledges support from the Mexican foundation Conacyt.

{}

\end{document}